\documentclass[superscriptaddress,twocolumn,pre]{revtex4}

\usepackage{amsmath}
\usepackage{graphicx,color}

\newcommand{\be}{\begin{equation}}
\newcommand{\ee}{\end{equation}}
\newcommand{\bea}{\begin{eqnarray}}
\newcommand{\eea}{\end{eqnarray}}

\newcommand{\erf}{\textrm{erf}}

\begin{document}
\sloppy

%-title page-%

\title{Mass-radius relation of Newtonian self-gravitating Bose-Einstein condensates \\
with short-range interactions: I. Analytical results}

\author{Pierre-Henri Chavanis}
%\email{chavanis@irsamc.ups-tlse.fr}
\affiliation{Laboratoire de Physique Th\'eorique (IRSAMC), CNRS and UPS,  Universit\'e de
Toulouse, France}

\begin{abstract}

We provide an approximate analytical expression of the mass-radius relation of a Newtonian self-gravitating Bose-Einstein condensate (BEC) with short-range interactions described by the Gross-Pitaevskii-Poisson system. These equations model astrophysical objects such as boson stars and, presumably, dark matter galactic halos. Our study connects the non-interacting case studied by Ruffini \& Bonazzola (1969) to the Thomas-Fermi limit studied by B\"ohmer \& Harko (2007). For repulsive short-range interactions (positive scattering lengths), there exists configurations of arbitrary mass but their radius is always larger than a minimum value. For attractive short-range interactions (negative scattering lengths), equilibrium configurations only exist below a maximum mass. Above that mass, the system is expected to collapse and form a black hole.  We also study the radius versus scattering length relation for a given mass. We find that equilibrium configurations only exist above a (negative) minimum scattering length. Our approximate analytical solution, based on a Gaussian ansatz, provides a very good agreement with the exact solution obtained by numerically solving a nonlinear differential equation representing hydrostatic equilibrium. Our analytical treatment is, however, easier to handle and permits to study the stability problem, and derive an expression of the pulsation period, by developing an analogy with a simple mechanical problem.
\end{abstract}

\maketitle

%-section: Introduction-%

\section{Introduction: a brief review}

Quantum mechanics predicts that the particles can be classified in two groups: fermions \cite{fermi,dirac} and bosons \cite{bose,einstein}. In early works, these systems were studied in the absence of interaction and fundamental concepts such as the Pauli exclusion principle for fermions  and the Bose-Einstein condensation for bosons were evidenced.
Fermi statistics forces the particles to occupy successively higher energy levels while bosons at zero temperature all condense into the ground state. In these ideal systems, the distribution of particles is spatially homogeneous. However, for very massive systems, as in astrophysical situations, one must take self-gravity into account. In that case, the system becomes spatially inhomogeneous and centrally condensed. This leads to the concepts of fermion \cite{chandra,shapiro} and boson \cite{liddle,leepang,jetzer,mielke} stars.

Soon after the discovery of the quantum statistics for fermions by Fermi \cite{fermi}  (1926) and Dirac \cite{dirac}  (1926), Fowler \cite{fowler} (1926), in a paper entitled ``Dense Matter'', understood that white dwarf stars owe their stability to the quantum pressure of the degenerate electron gas \footnote{The observations of Adams (1915) \cite{adams} had led to the discovery that white dwarf stars  such as Sirius B were very compact objects. As noted by Eddington (1926) \cite{eddington}: ``We have a star of mass about equal to the sun, and of radius much less than Uranus.'' These results could not be explained by classical physics.}. Therefore, gravitational collapse is avoided by Pauli's exclusion principle \footnote{As noted by Fermi (1926) \cite{fermi}, the Pauli (1925) \cite{pauli} principle was also formulated by Stoner (1924) \cite{stonerexclusion}.}. Fowler modeled a white dwarf star at zero temperature by a completely degenerate Fermi gas in gravitational equilibrium. He noted that the star is ``strictly analogous to one gigantic molecule in its lowest quantum state'' \footnote{This analogy takes even more sense when one realizes that the polytropic model of white dwarf stars that results from these considerations \cite{milne,chandra0} is mathematically similar to the Thomas-Fermi model of heavy atoms developed at the same period \cite{thomas,fermiatom}.}. As shown by Milne (1930) \cite{milne} and Chandrasekhar (1931a) \cite{chandra0}, the resulting structure is equivalent to a polytrope of index $n=3/2$ so that the mass-radius relation of classical white dwarf stars is $MR^3= 91.9 \hbar^6/m_e^3 G^3 (\mu H)^5$ where $m_e$ is the mass of the electron, $H$ the mass of the proton and $\mu$ the molecular weight.  Chandrasekhar (1931b,1935) \cite{chandra1,chandra2} took special relativity into account and showed that the star becomes equivalent to a polytrope of index $n=3$ in the ultra-relativistic limit. From that result, he deduced the existence of a maximum  mass $M_{Ch}= 3.09 (\hbar c/G)^{3/2}(\mu H)^{-2}$ above which there is no hydrostatic equilibrium \footnote{For a brief history of the discovery of the limiting mass of relativistic white dwarf stars, and the important contributions of Anderson (1929) \cite{anderson}, Stoner (1930) \cite{stoner} and Landau (1932) \cite{landau}, some of them having largely been forgotten, see Refs. \cite{blackman,prd,nono}.}. This mass $M_{Ch}= 5.75 M_{\odot}/\mu^2$ is of the order of the solar mass  ($M_{Ch}= 1.44 M_{\odot}$ for $\mu=2$) \footnote{As we approach the maximum mass, general relativity effects must be taken into account. They induce a dynamical instability when the radius becomes smaller than $1.0267\, 10^3 {\rm km}$ \cite{kaplan,chandratooper} which is the order of magnitude of the white dwarfs ($R\sim 5\, 10^3 {\rm km}$). The corresponding mass and mean density are $M= 1.4176M_\odot$ and $\rho= 6.220\, 10^{8} {\rm g \, cm}^{-3}$.}.  Massive stars cannot pass into the white dwarf stage. They undergo gravitational collapse and become neutron stars in which gravitational contraction is arrested by the quantum pressure of the neutrons \footnote{Two years after Chadwick's discovery of the neutron in 1932,  Baade \& Zwicky (1934) \cite{baade} proposed the idea of neutron stars and suggested that they would be formed in supernova explosions. The possibility of cold, dense stars composed principally of neutrons was also contemplated by Landau (1938) \cite{landau2}.}. The structure of neutron stars was studied by Oppenheimer \& Volkoff  (1939) \cite{ov} using general relativity. They obtained a maximum mass $M= 0.376(\hbar c/G)^{3/2}m_n^{-2}= 0.7 M_\odot$ (and a corresponding radius $R= 9.6 \, {\rm km}$ and density $\rho= 5\, 10^{15} {\rm g\,  cm}^{-3}$) above which no equilibrium state exists (the Chandrasekhar calculations, neglecting general relativity, lead to a larger value of the maximum mass $M= 5.75 M_{\odot}$). In that case, nothing prevents the gravitational contraction of the star which becomes a black hole. The mass-radius relation $M(R)$ of neutron stars makes a spiral while the mass-central density relation $M(\rho_0)$ presents damped oscillations. The series of equilibria becomes unstable after the first mass peak which corresponds to the maximum mass \cite{ov,mz,htww,chavrelat}. The existence of a limiting mass for self-gravitating fermions, the Chandrasekhar mass,  is striking because it can be expressed in terms of fundamental constants similarly to Bohr's radius of the atom \cite{nobel}. This limiting mass arises because of relativistic effects. For a non-relativistic fermion star, we have the mass-radius relation $MR^3= 91.9 \hbar^6/m^8 G^3$ that is valid for small masses $M\ll M_{Ch}$. For general relativistic fermion stars \cite{ov}, the maximum mass and the minimum radius can be written  $M_{Ch}= 0.376 M_P^3/m^2$ and $R= 3.52 (M_P/m)^2 l_P$  where $M_P=(\hbar c/G)^{1/2}$ is the Planck mass and $l_P= (\hbar G/c^3)^{1/2}$ is the Planck length. The radius can be expressed in terms of the Compton wavelength of the fermions $\lambda_c=\hbar/mc$ as  $R= 3.52(M_P/m)\lambda_c$. For $m\sim 1 {\rm GeV}/c^2$, we get $M_{Ch}\sim 10^{30}{\rm kg}\sim M_{\odot}$ and $R\sim {\rm km}$.

The preceding results are valid at $T=0$. The self-gravitating Fermi gas at non-zero temperature has been studied in the context of statistical mechanics by Hertel \& Thirring (1971) \cite{ht},  Messer (1981) \cite{messer} and more recently by Chavanis (2002) \cite{pre} (an exhaustive description of the phase diagrams in canonical and microcanonical ensembles is presented in the review \cite{ijmpb}). In these studies, the system must be enclosed within an artificial ``box'' so as to avoid its complete evaporation. For large systems, one recovers the classical isothermal self-gravitating gas \cite{chandra} that may undergo a gravothermal catastrophe \cite{gc} in the microcanonical ensemble (fixed energy) or an isothermal collapse \cite{ic} in the canonical ensemble (fixed temperature). If the particles are fermions, the gravitational collapse stops when quantum degeneracy effects (Pauli exclusion principle) come into play. This leads to the formation of a composite structure made of a completely degenerate and very compact nucleus (fermion ball) surrounded by a dilute atmosphere \cite{ijmpb}. Self-gravitating fermions were also discussed in cosmology, in models where dark matter is made of massive neutrinos \footnote{Self-gravitating degenerate neutrino stars were initially suggested as a model of quasars \cite{markov}.}.  Originally, the self-gravitating Fermi gas with  neutrino masses in the $\sim {\rm eV}/c^2$  range  was proposed by Ruffini and co-workers (1982-1990) \cite{fabbri,stella,gao} as a model for dark matter halos ($R\sim 10^{23}{\rm cm}\sim 100{\rm kpc}$ and $M\sim 10^{45}{\rm g}\sim 10^{12}M_{\odot}$) and clusters of galaxies. Then, Viollier and coworkers (1997-2001) \cite{bilic1,bilic2,bilic3} suggested that degenerate superstars composed of weakly interacting fermions in the $\sim 10{\rm keV}/c^2$ range could be an alternative to the supermassive black holes  that are reported to exist at the centers of galaxies (e.g. $M= 2.6 \, 10^{6} M_{\odot}$ and $R= 18 {\rm mpc}$ in our Galaxy). Finally, Bilic {\it et al.} (2003)  \cite{bilic4} showed that a weakly interacting fermionic gas at finite temperature could  provide a self-consistent model of dark matter that describes both the center and the halo of the galaxies. Since the density of a self-gravitating isothermal sphere decreases as $r^{-2}$ at large distances, this model is consistent with the flat rotation curves of the galaxies. On the other hand, since the core is degenerate in the sense of quantum mechanics (Pauli exclusion principle), it leads to flat density profiles and  avoids the cusp problem of cold dark matter (CDM) models (see below). In addition, the gravitational collapse of fermionic matter leads to a compact object (fermion ball) at the center of the galaxy  that could mimic a central black hole.

Coincidentally, the Fermi-Dirac statistics also arises in the theory of violent relaxation developed by Lynden-Bell (1967) \cite{lb}  for the Vlasov-Poisson system.  However, in that case, the origin of the ``degeneracy'' is due to dynamical constraints (Liouville's theorem) instead of quantum mechanics (Pauli's principle). This theory was initially developed to  describe collisionless stellar systems such as elliptical galaxies. In that case, the non-degenerate limit may be the most relevant \cite{lb}. However, as proposed by Kull {\it et al.} (1996) \cite{kull} and Chavanis \& Sommeria (1998) \cite{csmnras}, this approach (with dynamical degeneracy retained) could also apply to dark matter halos  and  provide a much more  efficient relaxation mechanism than in the fermion scenario. Indeed, the violent relaxation of collisionless systems (leading to the Lynden-Bell statistics) takes place on a few dynamical times while the collisional relaxation of fermions (leading to the Fermi-Dirac statistics) is larger than the age of the universe by several orders of magnitude. Therefore, it is not clear how the fermions have thermalized and how they can possess sufficiently large temperatures. By contrast, the Lynden-Bell theory predicts a large effective temperature (even if $T=0$ initially), a $r^{-2}$ density profile at large distances  consistent with the flat rotation curves of galaxies and an effective exclusion principle at short distances that could avoid the cusp problem and lead to fermion balls mimicking black holes, just like in the fermionic scenario.

Therefore, fermion stars rapidly found applications in relation to white dwarf stars, neutron stars, massive neutrinos in dark matter models, and in the violent relaxation of collisionless self-gravitating systems.

The concept of boson stars  was introduced by Kaup (1968) \cite{kaup} and Ruffini \& Bonazzola (1969) \cite{rb} although no astrophysical application of these objects was known at that time. They were just hypothetical stars whose main interest was ruled by the fundamental  laws of physics that govern their structure. In a sense, boson stars are descendent of the so-called {\it geons} of Wheeler (1955) \cite{wheeler} except that they are built from scalar particles (spin-$0$) instead of electromagnetic fields, i.e. spin-$1$ bosons. Kaup and Ruffini \& Bonazzola considered the $T=0$ limit \footnote{The case of self-gravitating bosons at nonzero temperature has been studied in \cite{ingrosso,nikolic}.} at which the bosons form a Bose-Einstein condensate (BEC) and showed that the concept of an equation of state breaks down. Indeed, at zero temperature a boson gas has a vanishing pressure and is unable to sustain any configuration of equilibrium in the classical perfect fluid approximation. But if one considers a quantum system of massive self-gravitating bosons in their ground state self-consistently, stable equilibrium configurations exist. In that case, all the bosons are in the same quantum state and they are described by a unique wave function $\psi({\bf r})$. Therefore, boson stars can be regarded as {macroscopic quantum states} that are only prevented from collapsing gravitationally by the Heisenberg uncertainty principle $\Delta x\Delta p\sim \hbar$.   In the Newtonian approximation, a self-gravitating BEC is described by the Schr\"odinger-Poisson system and in the relativistic case one must couple the Klein-Gordon equation to the Einstein field equations. The Newtonian approximation is valid for sufficiently small masses and yields the mass-radius relation $MR= 9.9\hbar^2/Gm^2$ (where $R$ is the radius containing $99\%$ of the mass) \cite{rb}. The radius decreases as mass increases, like for classical white dwarf stars, but the scaling is different.  When relativistic effects are taken into account, there exists a maximum mass, the Kaup mass $M_{Kaup}= 0.633M_P^2/m$, above which no equilibrium configuration exists \cite{kaup,rb}. In that case, the system collapses to a black hole. The minimum radius  $R_{min}= 6.03\lambda_c$ corresponding to the Kaup mass is of the order of the Compton wavelength $\lambda_c=\hbar/mc$ of the bosons. These results were re-derived independently by Thirring (1983) \cite{thirring} and Breit {\it et al.} (1984) \cite{breit}.

There exists remarkable similarities between boson and neutron stars
\cite{kaup,rb,breit,seidel90}. For example, the mass-central density
relation $M(\rho_0)$ of boson stars exhibits damped oscillations and
the series of equilibria becomes dynamically unstable after the first
mass peak corresponding to the maximum mass
\cite{leestab,gleiser,kusmartsev}. On the other hand, the mass-radius
relation $M(R)$ has a snail-like (spiral) structure, and the
mass-particle number relation $M(N)$ presents cusps.  The Newtonian
approximation is valid for sufficiently low densities corresponding to
$M\ll M_{Kaup}$ and $R\gg \lambda_c$. On the other hand, boson stars
become relativistic when they approach the maximum mass $M_{Kaup}=
0.633M_P^2/m$ and the corresponding radius $R_{min}=
6.03\lambda_c$. Except for matters of scales, all these results are
remarkably similar to those obtained for neutron stars
\cite{ov,mz,htww,chavrelat}. However, there also exists crucial
differences between self-gravitating objects made of fermions or
bosons. In particular, boson stars are stopped from collapsing by
Heisenberg's uncertainty principle while for fermion stars
gravitational collapse is avoided by Pauli's exclusion principle. This
difference is reflected in the critical mass of stable configurations:
$M_{Ch}\sim M_P^3/m^2$ for fermions and $M_{Kaup}\sim M_P^2/m$ for
bosons. This difference is due principally to the fact that all the
$N$ bosons are in the ground state whereas the $N$ fermions, according
to the Pauli exclusion principle, are distributed in the $N$ lowest
energy states of the phase space. As a result, the mass of boson stars
$M_{Kaup}\sim M_P^2/m$ is generally much smaller than the mass of
fermion stars $M_{Ch}\sim M_P^3/m^2$. They differ by a factor
$m/M_P\ll 1$. For example, for $m\sim 1 {\rm GeV}/c^2$ for which
$m/M_P\sim 10^{-19}$, one can estimate the total mass of a boson star
to be $M\sim 10^{11}{\rm kg}\sim 10^{-19}M_{\odot}$ and its radius
$R\sim 10^{-15}{\rm m}$ yielding a density $10^{38}$ times that of a
neutron star.  Contrary to the mass of neutron stars that is of the
order of a solar mass, the mass of these mini boson stars is too small
to be astrophysically relevant. They could play a
role, however, if they exist in the universe in large quantity or if
the mass $m$ of the bosons is extraordinary small leading to
macroscopic objects with a mass comparable to the mass of the sun (or
even larger) \cite{mielke}. This is the case, in particular, for
axionic boson stars that could account for the mass of MACHOs (between
$0.3$ and $0.8$ $M_{\odot}$) if the axions have a mass $m\sim 10^{-10}{\rm
eV}/c^2$ \cite{mielkeschunck}.

Colpi {\it et al.} (1986) \cite{colpi} considered the case of self-interacting scalar fields with a $\frac{1}{4}\lambda |\phi|^4$ term and found that the resulting configurations differ markedly from the non-interacting case even when $\lambda\ll 1$. In that case, the maximum mass $M= 0.06 \sqrt{\lambda}M_P^3/m^2$ may be comparable with the Chandrasekhar mass of self-gravitating fermions when $\lambda\sim 1$. Similarly, the radius $R\sim \sqrt{\lambda}(M_P/m)\lambda_c$ of self-interacting boson stars may be much larger than the Compton wavelength and become comparable with the radius of fermion stars. For $m\sim 1{\rm GeV}/c^2$, boson star configurations exist with a mass  $M\sim 10^{30}\sqrt{\lambda}{\rm kg}$ and a radius $R\sim \sqrt{\lambda} {\rm km}$ similar to those of neutron stars. For smaller masses $m\sim 1 {\rm MeV}/c^2$, we get $M\sim 10^{36}\sqrt{\lambda} {\rm kg}$ and $R\sim 10^6 \sqrt{\lambda} {\rm km}$. The radius is comparable to that of the sun but it encloses $10^6$ solar masses. These parameters are reminiscent of supermassive black holes in Active Galactic Nuclei, so that boson stars, like fermion stars, could be an alternative to black holes \cite{schunckliddle}. These enhancements of mass and size are due to the parameter $\sqrt{\lambda} M_P/m$ that is very large  even for small $\lambda$ due to the smallness of $m$ relative to $M_P$.  Therefore, self-coupling can significantly change the physical dimensions of boson stars, making them much more astrophysically interesting. The self-interaction has the same effect on the bosons as the exclusion principle on fermions. It plays the role of an interparticle {\it repulsion} (for $\lambda>0$) that  dominates over uncertainty pressure and prevents catastrophic gravitational collapse. Colpi {\it et al.} \cite{colpi} showed that a quartic self-interaction is equivalent to an effective  barotropic pressure.  At low densities $p=(\lambda\hbar^3/4m^4c^5)\epsilon^2$ (polytrope $n=1$) and at high densities $p=\epsilon/3$ (linear) like in the core of neutron stars \cite{ov,mz,htww,chavrelat} (this result is not explicitly given in \cite{colpi} but it can be obtained as a limit of their Eq. (17)). This strengthens the analogy between boson stars and neutron stars. Since boson stars may have masses comparable to the mass of neutron stars, or even larger, they become astrophysical relevant and may play a role in the problem of dark matter \cite{colpi,mielkeschunck}.

A wide ``zoology'' of exotic particles that could form dark matter has been proposed. In particular, many grand unified theories in particle physics predict the existence of various exotic  bosons (e.g. axions, scalar neutrinos, neutralinos) that should be present in considerable abundance in the universe and comprise (part of) the cosmological missing mass \cite{primack,overduin}. Although the bosonic particles have never been detected in accelerator experiments, they are considered as leading candidates of dark matter and might play a significant role in the evolution and the structure of the universe. The formation of boson stars was investigated by Madsen \& Liddle (1990) \cite{madsen} and is now relatively well understood. A spatially homogeneous distribution of self-gravitating bosons can undergo a sort of Jeans instability as described  by Khlopov {\it et al.} (1985) \cite{khlopov} and Bianchi {\it et al.} (1990) \cite{bianchi}. When the perturbation has sufficiently grown, the cloud collapses under its own gravity at first in free fall. Then, as nonlinear gravitational effects become important at higher densities, the configuration  starts to oscillate and  settles into a compact bosonic object  through the radiation of the scalar field. This gravitational cooling process has been evidenced and studied by Seidel \& Suen (1994) \cite{seidel94}. This is a dissipationless mechanism similar in some respect to the violent relaxation of collisionless stellar systems \cite{lb,albada} but ending on a unique final state (boson star) independent on the initial conditions. Therefore, (mini) boson stars could be the constituents of dark matter halos \cite{colpi,mielkeschunck}. Later, it was suggested that dark matter halos themselves could be gigantic self-gravitating BECs. This idea was introduced in order to solve basic  problems inherent to cold dark matter models.

Dark matter is one of the most important puzzles in modern physics and cosmology. Cold dark matter ($\Lambda$CDM) models with a small cosmological constant comprising weakly interacting massive particles (WIMPs), such as the lightest neutralinos, are presently favored by theorists over hot dark matter (HDM) models comprising relativistic light neutrinos. Although the cold dark matter (CDM) model is popular and remarkably successful in explaining the large-scale structure of the universe \cite{ratra}, it seems to encounter many problems on the scale of galactic or sub-galactic structures. Indeed, CDM simulations lead to $r^{-1}$ cuspy density profiles at galactic centers (in the scales of the order of $1$ kpc and smaller) \cite{nfw} while most rotation curves indicate a smooth core density \cite{observations}. On the other hand, the predicted number of satellite galaxies around each galactic halo is far beyond what we see around the Milky Way \cite{satellites}. These problems might be solved, without altering the virtues of CDM models, if the dark matter is composed of scalar particles in a cold BEC. The wave properties of the dark matter may stabilize the system against gravitational collapse as a consequence of the Heisenberg uncertainty principle, providing halo cores instead of cuspy profiles. The resulting  coherent configuration may be understood as the ground state of some gigantic bosonic atom where the ultra-light boson particles are condensed in a single macroscopic quantum state $\psi({\bf r})$. In these models, the formation of dark matter structures at small scales is suppressed by the quantum uncertainty principle.  This property could alleviate the problems of the CDM model such as the cusp problem and the missing satellite problem.

The first suggestion that galactic halos are formed by bosons,  either in their quantum ground state (BEC)  or in an appropriate isothermal distribution, is due to Baldeschi {\it et al.} (1983) \cite{baldeschi}. In order to yield masses and sizes that agree with those of
galactic halos ($M\sim 10^{45}\, {\rm g}$  and $R\sim 10^{23}\, {\rm cm}$), the mass of the bosons must be extremely small $m\sim 10^{-24}\, {\rm eV}/c^2$ (estimated with the Newtonian mass-radius relation). The idea that galactic halos could be a giant system of ``Bose liquid'' was also advanced by Sin (1994) \cite{sin} who studied rotation curves induced by self-gravitating BECs.   Sin  considered the ultralight pseudo Nambu-Goldstone boson (axion) appearing in the late-time cosmological phase transition theories as a major dark matter candidate. Since it is almost massless $m\sim 10^{-24}\, {\rm eV}/c^2$, its nature is more wavelike than particlelike so that the dark matter distribution must be treated quantum mechanically. Therefore, galactic halos can be considered as self-gravitating Bose liquids whose collapse is prevented by the uncertainty principle. At the galactic scale, the Newtonian approximation turns out to be relevant. Sin considered condensation wave functions for the galactic halo that have nodes (excited states) because the rotation curve of the zero node solution falls too fast to explain the flatness of the rotation curve of many galaxies.  Because of the systems' wavelike nature, he found that the rotation curve of galaxies has a ripplelike fine structure that seems to agree with observations. However, this may be a coincidence because the presence of wiggles in the rotation curves is rather related to spiral arms inside the disks, and excited states are generally unstable \cite{leestab}. Indeed, the excited modes decay to the ground state through emission of gravitational radiation, a process similar to atomic transitions \cite{ferrell,balakrishna}. In this sense, boson stars are like gravitational atoms.

An alternative scalar field
matter model for dark halos of galaxies was developed by Schunck
(1998) \cite{schunckpreprint}. He obtains rotation curves
that give a very good agreement with observations of spiral and dwarf
galaxies.  In this model, the density decreases at large distances
like $r^{-2}$ yielding asymptotically flat rotation
curves. Furthermore, this model produces oscillations around this
asymptotic value that can match, in some cases, the data. Since these
oscillations (``wiggles'') do not correspond to excited states, the
solutions are stable. Although these results are
obtained in the Newtonian limit, it is shown that the radial pressure
coming from general relativity plays a role for the rotation velocity
that is comparable to the contribution of the normal part. On the other
hand, in the strongly relativistic case, this model leads to massive
objects with large redshift values and rotation velocities. This might
be an explanation for the large energy contributions seen in quasars
\cite{schunckpreprint}. At about the same period, Matos \& Guzm\'an (1999)
\cite{matosguzman} introduced a model in which dark matter is a scalar
field embedded in an exponential (and later cosh) scalar potential.
This model produces a density profile of the form $1/(r^2+r_c^2)$
which gives a good agreement with the rotation curve of galaxies and
accounts for their flatness at large distances \cite{guzmanmatos}. It
also explains the suppression of subgalactic structures (since it
produces a sharp cut-off in the mass power spectrum) and the
smoothness of galaxy core halos \cite{mu}. This model has only one
free parameter, the scalar field mass, whose determined value $m\sim
10^{-23}{\rm eV}/c^2$ accounts both for the typical mass of galactic
halos and for cosmological observations (two {\it a priori}
independent measurments) \cite{matosall}. A similar mass was found by
Arbey {\it et al.} (2001) \cite{arbey1} and Silverman \& Mallett
(2002) \cite{silverman}.

In these models, the self-interaction of the particles is neglected and the mass of the bosons must be extremely small in order to reproduce the characteristic mass and size of galactic halos. Such an ultralight scalar field (e.g. an axion) with $m\sim 10^{-24}\, {\rm eV}/c^2$ was called ``fuzzy cold dark matter'' (FCDM) by Hu {\it et al.} (2000)  \cite{hu} who discussed its overall cosmological behavior. An alternative to this unnatural small mass is to take self-coupling into account. Indeed, in the context of boson stars, the work of Colpi {\it et al.} (1986) \cite{colpi} has demonstrated  that, for the same value of the boson mass $m$, even a small coupling can considerably change the mass and size of self-gravitating BECs. In that case, values of the mass in the ballpark of an ${\rm eV}/c^2$ may be compatible with a size of a few kiloparsecs. Lee \& Koh (1996) \cite{leekoh} investigated relativistic boson stars with a self-interacting scalar field as a model of galactic halos. A massive scalar field or a boson condensate with quartic - or close to quartic - self-coupling was also proposed as a possible dark matter candidate by Peebles (2000) \cite{peebles}, who called it ``fluid dark matter'' and by Goodman (2000) \cite{goodman} who called it ``repulsive dark matter''.  Similarly, Arbey {\it et al.} (2003) \cite{arbey} considered a self-coupled charged scalar field which is equivalent to a self-gravitating Bose condensate. For $m^4/\lambda\sim 50\, ({\rm eV}/c^2)^4$, they obtained a very good agreement with the measurements of the circular speed of the dwarf spiral DDO154. B\"ohmer \& Harko (2007) \cite{bohmer} pursued the idea that dark matter is in the form of a self-gravitating BEC and studied the condensate by using the non-relativistic Gross-Pitaevskii equation coupled to the Poisson equation. They took  self-interaction into account via a quartic nonlinearity and used the Thomas-Fermi approximation which becomes exact for $N\rightarrow +\infty$. Under these assumptions, the BEC is equivalent to a barotropic fluid with a polytropic equation of state of index $n=1$  \cite{leekoh,goodman,arbey,bohmer}. This leads to a length-scale $R=\pi (a\hbar^2/Gm^3)^{1/2}$ for bound objects that is independent on their mass ($a$ is the scattering length). B\"ohmer \& Harko  considered the example of a galactic dark matter halo extending up to $R= 10\ {\rm kpc}= 3.08 \, 10^{22}\, {\rm cm}$ with a mass of the order of $M= 3\, 10^{11} M_{\odot}$ yielding an average density $\rho= 5.30\, 10^{-24}\, {\rm g \, cm}^{-3}$. For $a=5.77 \, 10^{-7} {\rm cm}$ (a typical value in terrestrial BEC experiments \cite{revuebec}) they found that the mass of the particles forming the condensate dark matter halo is of the order of the eV (more precisely,  $m=1.44 \, {\rm eV}/c^2$ for $a=10^6 \, {\rm fm}$ and $m=14 \, {\rm meV}/c^2$ for $a=1 \, {\rm fm}$). They also determined the rotation curves created by a BEC and found a very good agreement with the observational data for several low surface brightness galaxies. For recent studies of scalar field/BEC dark matter models see, e.g., Refs. \cite{bmn,fmt,mvm,fm,lee09,lee,briscese,sm}.

This detailed introduction shows that the idea of boson stars has a
long and rich history and that they may play a role in different areas
of astrophysics. If we consider dark matter halos, the Newtonian
approximation is sufficient. In previous works, two limits have been
considered. Ruffini \& Bonazzola \cite{rb} neglect the
self-interaction of the particles ($a=0$) and solve the
Schr\"odinger-Poisson system. In that case, the equilibrium state
results from the balance between the gravitational attraction and the
Heisenberg principle equivalent to a quantum pressure. Alternatively,
B\"ohmer \& Harko \cite{bohmer} take into account the self-interaction
of the particles ($a>0$) and study the Gross-Pitaevskii-Poisson (GPP)
system. They consider the Thomas-Fermi (TF) limit which amounts to
neglecting the quantum pressure. In that case, the equilibrium state
results from the balance between the gravitational attraction and the
small-scale repulsion due to scattering. As we shall see, the
TF limit is valid if $GN^2m^3a/\hbar^2\gg 1$. In the present
work, we shall connect these two limits by considering the general
case where both short-range interactions and quantum pressure are
taken into account. We shall also treat the case where the
self-interaction is attractive instead of repulsive. Since atoms may
have negative scattering lengths in terrestrial BEC experiments
\cite{revuebec}, it may be useful to consider the possibility of
attractive interactions in our general study. The TF
approximation cannot be employed in that case and we have to use the
complete set of equations. Attractive self-interaction is equivalent
to a negative pressure $p=-|k|\rho^2$ that adds to the gravitational
attraction. In that case, we find the existence of a maximum mass
$M_{max}= 1.012\hbar/\sqrt{|a|Gm}$ above which the system cannot be in
equilibrium. This maximum mass can be very small, as small as the
Planck mass $M_P$ (!), meaning that when the self-interaction is
attractive the system is very unstable. When applied to a cosmological
context \cite{cosmobec}, an attractive self-interaction could enhance
the gravitational collapse and accelerate the formation of structures
in the universe.

%At the same time, since it yields an equation of
%state with a negative pressure (which is allowed for a BEC), an
%attractive self-interaction may lead to an acceleration of the
%expansion of the universe in the early times when the density is
%high \cite{cosmobec}.

The paper is organized as follows. In Secs. \ref{sec_gpp} and \ref{sec_exact} we provide general results concerning the Gross-Pitaevskii-Poisson system. We specifically consider the non-interacting case and the Thomas-Fermi limit. In Sec. \ref{sec_ansatz}, we obtain an analytical approximate expression of the mass-radius relation of self-gravitating BECs with positive or negative scattering lengths by using a Gaussian ansatz for the wave function and developing a simple mechanical analogy.  In Sec. \ref{sec_jeans}, we study the Jeans instability of an infinite homogeneous self-gravitating BEC by taking into account  the self-interaction of the particles that was ignored in previous works. In paper II \cite{paper2}, we show that our approximate analytical approach gives a good agreement with the exact results obtained by numerically solving the equation of hydrostatic equilibrium. In Paper III, we extend our analytical approach to more general situations.

\section{The Gross-Pitaevskii-Poisson system}
\label{sec_gpp}

\subsection{The mean-field Gross-Pitaevskii equation}
\label{sec_mfgp}

Following B\"ohmer \& Harko \cite{bohmer}, we model dark matter halos as a self-gravitating Bose-Einstein condensate with short-range interactions. Since the cosmic BEC has a relatively low mean mass density, we can use the Newtonian approximation. At $T=0$, all the bosons have condensed \footnote{The condensation of bosons of mass $m$  takes place provided that the de Broglie wavelength $\lambda_{dB}=\sqrt{2\pi\hbar^2/mk_BT}$ exceeds the mean separation $n^{-1/3}$. The  critical condensation temperature is $k_B T_c= 2\pi \hbar^2 n^{2/3}/(\zeta(3/2))^{2/3}m$. If we assume an adiabatic cosmological expansion of the universe, the temperature has the same dependance $T\propto n^{2/3}$ on the number density of the particles. This implies that Bose-Einstein condensation occurs if the mass of the particles satisfies the condition $m<1.87\, {\rm eV}/c^2$ \cite{fmt,bhcosmo}.}  and the system is described by one order parameter $\psi({\bf r},t)$ called the condensate wave function.
 In the mean-field approximation, the ground state properties of the condensate are described by the Gross-Pitaevskii equation \cite{gross,pitaevskii}:
\begin{equation}
\label{mfgp1}
i\hbar \frac{\partial\psi}{\partial t}({\bf r},t)=-\frac{\hbar^2}{2m}\Delta\psi({\bf r},t)+m\Phi_{tot}({\bf r},t)\psi({\bf r},t),
\end{equation}
\begin{eqnarray}
\label{mfgp2}
\Phi_{tot}({\bf r},t)=\int \rho({\bf r}',t) u(|{\bf r}-{\bf r}'|)\, d{\bf r}',
\end{eqnarray}
\begin{eqnarray}
\label{mfgp3}
\rho({\bf r},t)=Nm|\psi({\bf r},t)|^2,
\end{eqnarray}
\begin{eqnarray}
\label{mfgp4}
\int |\psi({\bf r},t)|^2\, d{\bf r}=1.
\end{eqnarray}
Equation (\ref{mfgp4}) is the normalization condition, Eq. (\ref{mfgp3}) gives the density of the BEC, Eq. (\ref{mfgp2}) determines the associated potential and Eq. (\ref{mfgp1}) determines the wave function. We assume that the potential of interaction can be written as $u=u_{LR}+u_{SR}$ where $u_{LR}$ refers to the long-range gravitational interaction and $u_{SR}$ to the short-range interaction. We assume that the short-range interaction corresponds to binary collisions that can be modeled by the pair contact potential $u_{SR}({\bf r}-{\bf r}')=g\delta({\bf r}-{\bf r}')$, where the coupling constant (or pseudo-potential) $g$ is related to the $s$-wave scattering length $a$ through $g=4\pi a\hbar^2/m^3$ \cite{revuebec}. For the sake of generality, we allow $a$ to be positive or negative ($a>0$ corresponds to short-range repulsion and $a<0$ corresponds to short-range attraction). Under these conditions, the total potential can be written $\Phi_{tot}=\Phi+h(\rho)$ where $\Phi$ is the gravitational potential and
\begin{equation}
\label{mfgp5}
h(\rho)=g\rho=gNm|\psi|^2,
\end{equation}
is an effective potential modeling short-range interactions. When this form of potential is substituted in Eq. (\ref{mfgp2}), we obtain the Gross-Pitaevskii-Poisson (GPP) system
\begin{equation}
\label{mfgp6}
i\hbar \frac{\partial\psi}{\partial t}=-\frac{\hbar^2}{2m}\Delta\psi+m(\Phi+h(\rho))\psi,
\end{equation}
\begin{equation}
\label{mfgp7}
\Delta\Phi=4\pi G\rho=4\pi G Nm |\psi|^2,
\end{equation}
that will be the object of focus in this paper. In the general formalism developed in the sequel, we will consider an arbitrary potential $h(\rho)$. However, for specific applications, we will consider the potential (\ref{mfgp5}). More general situations will be studied in Paper III.

\subsection{The Madelung transformation}
\label{sec_mad}

Let us use the Madelung \cite{madelung} transformation to rewrite the GPP system in the form of hydrodynamic equations. From the wave function
\begin{equation}
\label{mad1}
\psi({\bf r},t)=A({\bf r},t) e^{iS({\bf r},t)/\hbar}
\end{equation}
where $A({\bf r},t)$ and $S({\bf r},t)$ are real functions, we introduce the density and velocity fields
\begin{equation}
\label{mad2}
\rho=Nm|\psi|^2=NmA^2, \qquad  {\bf u}=\frac{1}{m}\nabla S.
\end{equation}
We note that the flow defined in this way is irrotational since $\nabla\times {\bf u}={\bf 0}$. Substituting Eq. (\ref{mad1}) in Eq. (\ref{mfgp6}) and separating real and imaginary parts, we obtain
\begin{equation}
\label{mad3}
\frac{\partial\rho}{\partial t}+\nabla\cdot (\rho {\bf u})=0,
\end{equation}
\begin{equation}
\label{mad4}
\frac{\partial S}{\partial t}+\frac{1}{2m}(\nabla S)^2+m\Phi+m h(\rho)+Q=0,
\end{equation}
where
\begin{equation}
\label{mad5}
Q=-\frac{\hbar^2}{2m}\frac{\Delta \sqrt{\rho}}{\sqrt{\rho}}=-\frac{\hbar^2}{4m}\left\lbrack \frac{\Delta\rho}{\rho}-\frac{1}{2}\frac{(\nabla\rho)^2}{\rho^2}\right\rbrack,
\end{equation}
is the quantum potential. The first equation is similar to the equation of continuity in hydrodynamics. The second equation has a form similar to the classical Hamilton-Jacobi equation with an additional quantum term. It  can also be interpreted as a generalized Bernouilli equation for a potential flow.
 Taking the gradient of Eq. (\ref{mad4}) and using the well-known   identity $({\bf u}\cdot \nabla){\bf u}=\nabla ({{\bf u}^2}/{2})-{\bf u}\times (\nabla\times {\bf u})$ which reduces to $({\bf u}\cdot \nabla){\bf u}=\nabla ({{\bf u}^2}/{2})$ for an irrotational flow, we obtain an equation similar to the Euler equation with an additional quantum potential
\begin{equation}
\label{mad6}
\frac{\partial {\bf u}}{\partial t}+({\bf u}\cdot \nabla){\bf u}=-\nabla h-\nabla\Phi-\frac{1}{m}\nabla Q.
\end{equation}
This equation shows that the effective potential $h$ appearing in the GP equation can be
interpreted as an enthalpy in the hydrodynamic equations.  We can rewrite Eq. (\ref{mad6}) in the form
\begin{equation}
\label{mad7}
\frac{\partial {\bf u}}{\partial t}+({\bf u}\cdot \nabla){\bf u}=-\frac{1}{\rho}\nabla p-\nabla\Phi-\frac{1}{m}\nabla Q,
\end{equation}
where $p({\bf r},t)$ is a pressure.  Since $h({\bf r},t)=h\lbrack \rho({\bf r},t)\rbrack$, the pressure $p({\bf r},t)=p\lbrack \rho({\bf r},t)\rbrack$ is a function of the density  (the flow is barotropic). The equation of state $p(\rho)$ is determined by the potential $h(\rho)$ through the relation
\begin{equation}
\label{mad8}
h'(\rho)=\frac{p'(\rho)}{\rho}.
\end{equation}
This yields $p(\rho)=\rho h(\rho)-H(\rho)$ where $H$ is a primitive of
$h$ \footnote{We note that for a BEC at $T=0$ the pressure arising
in the hydrodynamic equation (\ref{mad7}) has a meaning different from
the pressure of a normal fluid at finite temperature. It arises
directly from the short-range interactions between particles
encapsulated in the effective potential $h(\rho)$ and is not due to
thermal motion (since $T=0$). As a result, the pressure $p$ can be
{\it negative}! This is the case, in particular, for a BEC described
by the equation of state (\ref{mad9}) when the scattering length $a$
is negative
\cite{revuebec}.}. In conclusion, the GPP system is {\it equivalent}
to the ``hydrodynamic'' equations (\ref{mad3}), (\ref{mad7}) and
(\ref{mfgp7}). We shall refer to these equations as the quantum
barotropic Euler equations.  In the classical limit $\hbar\rightarrow
0$, the quantum potential disappears and we recover the ordinary
barotropic Euler equations \cite{bt}.  For a potential of the form
(\ref{mfgp5}), the equation of state is
\begin{equation}
\label{mad9}
p=\frac{2\pi a\hbar^2}{m^3}\rho^{2}.
\end{equation}
This is equivalent to a polytropic equation of state
\begin{equation}
\label{mad10}
p=K\rho^{\gamma},\qquad \gamma=1+\frac{1}{n},
\end{equation}
with a polytropic constant  $K=2\pi a\hbar^2/m^3$ and a polytropic index $n=1$ (i.e. $\gamma=2$).
Inversely, the effective potential associated with the general polytropic equation of state (\ref{mad10})  is $h(\rho)=\lbrack K\gamma/(\gamma-1)\rbrack \rho^{\gamma-1}$.

{\it Remark:} the quantum potential (\ref{mad5}) first appeared in the work of Madelung \cite{madelung} and was rediscovered by Bohm \cite{bohm} (it is sometimes called ``the Bohm potential''). We note the identity
\begin{equation}
\label{mad11}
-\frac{1}{m}\nabla Q\equiv -\frac{1}{\rho}\partial_j P_{ij},
\end{equation}
where $P_{ij}$ is the quantum stress (or pressure) tensor
\begin{equation}
\label{mad12}
P_{ij}=-\frac{\hbar^2}{4m^2}\rho\, \partial_i\partial_j\ln\rho,
\end{equation}
or
\begin{equation}
\label{mad13}
P_{ij}=\frac{\hbar^2}{4m^2}\left (\frac{1}{\rho}\partial_i\rho\partial_j\rho-\delta_{ij}\Delta\rho\right ).
\end{equation}
This shows that the quantum potential is equivalent to an anisotropic pressure.

\subsection{The time-independent GP equation}
\label{sec_formal}

If we consider a wave function of the form
\begin{equation}
\label{tigp1}
\psi({\bf r},t)=A({\bf r})e^{-i\frac{Et}{\hbar}},
\end{equation}
we obtain the  time-independent GP equation
\begin{eqnarray}
\label{tigp2}
-\frac{\hbar^2}{2m}\Delta\psi({\bf r})+m(\Phi({\bf r})+h(\rho))\psi({\bf r})=E\psi({\bf r}),
\end{eqnarray}
where $\psi({\bf r})\equiv A({\bf r})$ is real and $\rho({\bf r})=Nm\psi^2({\bf r})$. Dividing Eq. (\ref{tigp2}) by $\psi({\bf r})$, we get
\begin{equation}
\label{tigp3}
m\Phi+mh(\rho)-\frac{\hbar^2}{2m}\frac{\Delta \sqrt{\rho}}{\sqrt{\rho}}=E,
\end{equation}
or, equivalently,
\begin{equation}
\label{tigp4}
m\Phi+mh(\rho)+Q=E.
\end{equation}
This relation can also be obtained from the quantum Hamilton-Jacobi equation (\ref{mad4}) by setting $S=-Et$. Combined with the Poisson equation (\ref{mfgp7}), we obtain an eigenvalue equation for the wave function $\psi({\bf r})$ where the eigenvalue $E$ is the energy. In the following, we shall be interested by the fundamental eigenmode corresponding to the smallest value of $E$. For this mode, the wave function $\psi(r)$ is spherically symmetric and  has no node so that the density profile decreases monotonically with the distance.

\subsection{Hydrostatic equilibrium}
\label{sec_he}

The time-independent solution (\ref{tigp3}) can also be obtained from the quantum barotropic Euler equation (\ref{mad7}) since it is equivalent to the GP equation. The steady state of the quantum barotropic Euler equation (\ref{mad7}), obtained by taking $\partial_t=0$ and ${\bf u}={\bf 0}$, satisfies
\begin{equation}
\label{he1}
\nabla p+\rho\nabla\Phi-\frac{\hbar^2\rho}{2m^2}\nabla \left (\frac{\Delta\sqrt{\rho}}{\sqrt{\rho}}\right )={\bf 0}.
\end{equation}
This generalizes the usual condition of hydrostatic equilibrium by incorporating the contribution of the quantum potential. Equation (\ref{he1}) describes the balance between the gravitational attraction, the repulsion due to the quantum potential and the repulsion (for $a>0$) or the attraction (for $a<0$) due to the short-range interaction (scattering). This equation is equivalent to Eq. (\ref{tigp3}). Indeed, integrating Eq. (\ref{he1}) using Eq. (\ref{mad8}), we obtain Eq. (\ref{tigp3}) where the eigenenergy  $E$ appears as a constant of integration. Combining Eq. (\ref{he1}) with the Poisson equation (\ref{mfgp7}), we obtain the fundamental equation of hydrostatic equilibrium with quantum effects
\begin{equation}
\label{he2}
-\nabla\cdot \left (\frac{\nabla p}{\rho}\right )+\frac{\hbar^2}{2m^2}\Delta \left (\frac{\Delta\sqrt{\rho}}{\sqrt{\rho}}\right )=4\pi G\rho.
\end{equation}
For an equation of state of the form (\ref{mad9}), it becomes
\begin{equation}
\label{he3}
-\frac{4\pi a \hbar^2}{m^3}\Delta\rho+\frac{\hbar^2}{2m^2}\Delta \left (\frac{\Delta\sqrt{\rho}}{\sqrt{\rho}}\right )=4\pi G\rho.
\end{equation}
There are two important limits that we discuss in the following.

\subsection{The non-interacting case}
\label{sec_nic}

The non-interacting case corresponds to $a=0$. This is the situation first considered by Ruffini \& Bonazzola \cite{rb} and revisited by Membrado {\it et al.} \cite{membrado} with a different method. In that case, the condition of hydrostatic equilibrium (\ref{he3}) reduces to
\begin{equation}
\label{nic1}
\frac{\hbar^2}{2m^2}\Delta \left (\frac{\Delta\sqrt{\rho}}{\sqrt{\rho}}\right )=4\pi G\rho.
\end{equation}
This corresponds to the balance between the gravitational attraction and the repulsion due to the quantum pressure arising from the Heisenberg uncertainty principle. This equation can be solved numerically to yield the density profile \cite{rb,membrado}. The density decays smoothly to infinity (see, e.g., Fig. 1 in \cite{membrado}) so that its support is not compact, contrary to fermion stars at $T=0$ \cite{chandra}. The radius of the configuration containing $99\%$ of the mass has been computed in \cite{membrado}. They obtained the value
\begin{equation}
\label{nic2}
R_{99}=9.9 \frac{\hbar^2}{GMm^2}.
\end{equation}
This radius is much smaller than the gravitational Bohr radius $a_B=\hbar^2/(Gm^3)$ by a factor $1/N\ll 1$.  They also found that the quantum kinetic energy $\Theta_Q$, the potential energy $W$, the total energy $E_{tot}$ and the  eigenenergy $E$ (see their definitions in Sec. \ref{sec_ef}) are given by
%$\Theta_Q=0.05426 {G^2M^3m^2}/{\hbar^2}$, $W=-0.10852 {G^2M^3m^2}/{\hbar^2}$, $E_{tot}=-0.05426 {G^2M^3m^2}/{\hbar^2}$, $E=-0.16278{G^2M^2m^3}/{\hbar^2}$.
\begin{equation}
\label{nic3}
\Theta_Q=0.05426 \frac{G^2M^3m^2}{\hbar^2},
\end{equation}
\begin{equation}
\label{nic4}
W=-0.10852 \frac{G^2M^3m^2}{\hbar^2},
\end{equation}
\begin{equation}
\label{nic5}
E_{tot}=-0.05426 \frac{G^2M^3m^2}{\hbar^2},
\end{equation}
\begin{equation}
\label{nic6}
E=-0.16278\frac{G^2M^2m^3}{\hbar^2}.
\end{equation}
Finally, they determined numerically the rotation curve
\begin{equation}
\label{nic7}
v_c(r)=\left \lbrack \frac{GM(r)}{r}\right \rbrack^{1/2},
\end{equation}
produced by a Newtonian self-gravitating BEC (see Fig. 2 in \cite{membrado}) and mentioned application to galactic dark matter halos. To our knowledge, Membrado {\it et al.} (1989) \cite{membrado} were among the first authors in the literature to propose that dark matter halos could be a condensate boson sphere and to compute the corresponding rotation curve. However, they did not take into account self-coupling which is necessary to obtain physically relevant results.

\subsection{The Thomas-Fermi approximation}
\label{sec_tf}

The Thomas-Fermi (TF) approximation amounts to neglecting the quantum potential in Eq. (\ref{he1}). This is the situation considered by B\"ohmer \& Harko \cite{bohmer}  in relation to dark matter halos with repulsive self-interaction $a>0$. In that case, Eq. (\ref{he1}) reduces to the usual  condition of hydrostatic equilibrium
\begin{equation}
\label{tf1}
\nabla p+\rho\nabla\Phi={\bf 0}.
\end{equation}
This corresponds to the balance between the gravitational attraction and the repulsion due to the short-range interaction.  Combined with the Poisson equation (\ref{mfgp7}), we obtain the fundamental equation of hydrostatic equilibrium
\begin{equation}
\label{tf2}
-\nabla\cdot \left (\frac{\nabla p}{\rho}\right )=4\pi G\rho.
\end{equation}
For an equation of state of the form (\ref{mad9}), it can be rewritten
\begin{equation}
\label{tf3}
\Delta\rho+\frac{Gm^3}{a\hbar^2}\rho=0.
\end{equation}
This equation, which is equivalent to the Lane-Emden equation for a polytrope of index $n=1$,  can be solved analytically \cite{chandra}. The density profile is given by the formula
\begin{equation}
\label{tf4}
\rho(r)=\frac{\rho_0 R}{\pi r}\sin\left (\frac{\pi r}{R}\right ),
\end{equation}
where $\rho_0$ is the central density  and
\begin{equation}
\label{tf5}
R=\pi\left (\frac{a\hbar^2}{Gm^3}\right )^{1/2},
\end{equation}
is the radius of the configuration at which the density vanishes (compact support). The radius of a polytrope $n=1$ is independent on the mass $M$ \cite{chandra}. These results have been derived by several authors in the context of self-gravitating BECs \cite{leekoh,goodman,arbey,bohmer} using different formalisms. The radius containing $99\%$ of the mass is given by $R_{99}= 0.95424211R$. The central density is determined by the mass according to
\begin{equation}
\label{tf6}
\rho_0=\frac{\pi M}{4R^3}=\frac{M}{4\pi^2}\left (\frac{Gm^3}{a\hbar^2}\right )^{3/2}.
\end{equation}
 Using the analytical expression (\ref{tf4}) of the density profile, we find that the moment of inertia (\ref{v2}) and the internal energy (\ref{ef8}) are given by
\begin{equation}
\label{tf7}
I=\left (1-\frac{6}{\pi^2}\right )MR^2=(\pi^2-6)\frac{a\hbar^2 M}{Gm^3},
\end{equation}
\begin{equation}
\label{tf8}
U=\frac{GM^2}{4R}=\frac{1}{4\pi}\frac{G^{3/2}m^{3/2}M^2}{a^{1/2}\hbar}.
\end{equation}
On the other hand, in the TF approximation, the steady state equation (\ref{tigp3}) reduces to
\begin{equation}
\label{tf9}
m\Phi+\frac{4\pi a\hbar^2}{m^2}\rho=E.
\end{equation}
It can be used to determine the eigenenergy $E$. Indeed, if we evaluate this relation at $r=R$ at which $\Phi=-GM/R$ and $\rho=0$, we find that
\begin{equation}
\label{tf10}
E=-\frac{GMm}{R}=-\frac{1}{\pi}\frac{G^{3/2}m^{5/2}M}{a^{1/2}\hbar}.
\end{equation}
Then, Eq. (\ref{tf9}) with Eqs. (\ref{tf4}) and (\ref{tf10}) determine the gravitational potential $\Phi(r)$. On the other hand, multiplying Eq. (\ref{tf9}) by $\rho$, integrating over the configuration, and using the expressions (\ref{tf8}) and (\ref{tf10}) of the internal energy $U$ and eigenenergy $E$, we find that the potential energy (\ref{ef9}) is given by
\begin{equation}
\label{tf11}
W=-\frac{3GM^2}{4R}=-\frac{3}{4\pi}\frac{G^{3/2}m^{3/2}M^2}{a^{1/2}\hbar}.
\end{equation}
The total energy is
\begin{equation}
\label{tf12}
E_{tot}=U+W=-\frac{GM^2}{2R}=-\frac{1}{2\pi}\frac{G^{3/2}m^{3/2}M^2}{a^{1/2}\hbar}.
\end{equation}
Finally, using Eq. (\ref{tf4}), the rotation curve (\ref{nic7}) has the analytical expression
\begin{equation}
\label{tf13}
v_c^2(r)=\frac{4G\rho_0 R^2}{\pi}\left \lbrack \frac{R}{\pi r}\sin\left (\frac{\pi r}{R}\right )-\cos\left (\frac{\pi r}{R}\right )\right \rbrack.
\end{equation}
For $r\rightarrow 0$, the velocity increases linearly with $r$ as for a uniform sphere with density $\rho_0$: $v_c(r)\sim (4\pi\rho_0 G/3)^{1/2} r$. For $r\ge R$, we recover the Keplerian law $v_c(r)=({GM}/{r})^{1/2}$. The rotation curves created by dark matter halos made of self-gravitating BECs in the TF limit have been studied by  Arbey {\it et al.} \cite{arbey} and B\"ohmer \& Harko \cite{bohmer} who showed that they provide a good agreement with measured  rotation curves of certain spiral galaxies.

\subsection{Dimensional analysis and validity of the TF approximation}
\label{sec_dim}

In the absence of short-range interaction, the structure of the self-gravitating BEC results from the balance between the gravitational attraction and the quantum pressure arising from the Heisenberg uncertainty principle. Using dimensional analysis in Eq. (\ref{he3}), i.e. $\hbar^2/m^2R^4\sim GM/R^3$, we obtain the length-scale
\begin{equation}
\label{dim1}
R_Q= \frac{\hbar^2}{GMm^2},
\end{equation}
which gives the typical size of a self-gravitating BEC with mass $M$ without short-range interaction ($a=0$).

In the TF approximation, in which the quantum potential is negligible, the structure of the self-gravitating BEC results from the balance between the gravitational attraction and the short-range repulsion due to scattering (when $a>0$). Using dimensional analysis in Eq. (\ref{he3}), i.e. $(a\hbar^2/m^3R^2)(M/R^3)\sim GM/R^3$, we obtain the length-scale
\begin{equation}
\label{dim2}
R_a=\left (\frac{a\hbar^2}{Gm^3}\right )^{1/2},
\end{equation}
which gives the typical size of a self-gravitating BEC with scattering length $a>0$ in the TF approximation.

Considering Eq. (\ref{he3}) again, the quantum pressure  and the pressure arising from the short-range interaction become comparable when $(a\hbar^2/m^3R^2)(M/R^3)\sim \hbar^2/m^2R^4$, i.e. $Na/R\sim 1$. Estimating $R$ by Eq. (\ref{dim1}) or (\ref{dim2}), this condition can be rewritten $\chi\sim 1$ where we have introduced the important dimensionless parameter
\begin{equation}
\label{dim3}
\chi\equiv \frac{GN^2m^3a}{\hbar^2}.
\end{equation}
For $a>0$ and $\chi\gg 1$, we are in the TF limit in which the quantum potential is negligible. In that case, the equilibrium state results from a balance between repulsive scattering and gravitational attraction. Alternatively, for $\chi\ll 1$, we are in the non-interacting limit in which scattering is negligible. In that case, the equilibrium state results from a balance between quantum pressure and gravitational attraction. The transition between these two regimes occurs for $\chi\sim 1$. For a given value of the scattering length $a$, the TF limit is valid for $M\gg M_a$ where
\begin{equation}
\label{dim4}
M_a=\frac{\hbar}{\sqrt{Gma}},
\end{equation}
and the non-interacting limit is valid for $M\ll M_a$. For a given value of the mass $M$, the TF limit is valid for $a\gg a_Q$ where
\begin{equation}
\label{dim5}
a_Q=\frac{\hbar^2}{GM^2m},
\end{equation}
and the non-interacting limit is valid for $a\ll a_Q$ \footnote{For boson stars without self-interaction, the Hartree (mean-field) approximation, leading to the Schr\"odinger-Poisson system, is exact for $N\rightarrow +\infty$ (see \cite{yau} in the relativistic case). For self-coupled boson stars with fixed $a>0$, according to Eq. (\ref{dim3}), the TF approximation becomes exact for $N\rightarrow +\infty$. If we want to take into account the quantum potential,  we have to consider a nontrivial limit in which $N\rightarrow +\infty$ with  $a N^2$ fixed (this differs from the scaling $N\rightarrow +\infty$ with  $a N$ fixed considered by Lieb {\it et al.} \cite{lieb} in the case where the gravitational interaction is replaced by an external trapping potential). The Hartree (mean-field) approximation, leading to the Gross-Pitaevskii-Poisson system, becomes exact in that limit.}.

When $a>0$, the gravitational attraction is necessary to balance the repulsive quantum potential and the repulsive short-range interaction. When $a<0$, we could expect an equilibrium between the repulsive quantum potential and the attractive short-range attraction, in which gravitational effects are negligible. Using dimensional analysis in Eq. (\ref{he3}), i.e. $(|a|\hbar^2/m^3R^2)(M/R^3)\sim \hbar^2/m^2R^4$, we obtain the length-scale
\begin{equation}
\label{dim6}
R'_a\sim N |a|,
\end{equation}
which gives the typical radius of a non-gravitational BEC with attractive short-range interactions. However, as we shall see, such equilibria are unstable.

\subsection{Analogies between bosons and fermions}
\label{sec_analo}

The mean-field Gross-Pitaevskii equation (\ref{mfgp6}) also describes a gas of fermions  when one takes into account the quantum potential $Q$ arising from the Heisenberg uncertainty principle \footnote{Such a description assumes that the fermions have the same probability distributions  \cite{manfredi}. More fundamentally, the fermions should be described by a mixture of $N$ pure states, each of them having a wave function $\psi_i$ obeying the mean-field Schr\"odinger equation without nonlinearity (except the one due to the interaction) \cite{haas}.}. This description goes beyond the TF approximation and can be useful to regularize the dynamics at small scales \cite{bilic3}. In the case of fermions, we must also take into account the quantum pressure arising from the Pauli exclusion principle. It can be calculated from the Fermi-Dirac distribution function at $T=0$. In $d$ dimensions, the equation of state is $p=K\rho^{1+2/d}$ where $K=({1}/({d+2}))({d}/{2S_d})^{2/d}{(2\pi \hbar)^2}/{m^{2+2/d}}$ for spin $s=1/2$ fermions \cite{wdD}. This is equivalent to a polytrope of index $n=d/2$. This pressure term is the one that appears in the hydrodynamic equation (\ref{mad7}). Using Eq. (\ref{mad8}), it corresponds to an effective  potential of the form $h(\rho)=(d/2+1)K\rho^{2/d}$ in the Gross-Pitaevskii equation (\ref{mfgp6}). In particular, in $d=3$ dimensions, the pressure is $p=K\rho^{5/3}$ with $K=({1}/{5})({3}/{8\pi})^{2/3}{(2\pi \hbar)^2}/{m^{8/3}}$ leading to a potential $h=(5/2)K\rho^{2/3}$ and to a GP equation
\begin{equation}
\label{analo1}
i\hbar \frac{\partial\psi}{\partial t}=-\frac{\hbar^2}{2m}\Delta\psi+m\Phi\psi+\frac{(3\pi^2)^{2/3}}{2}N^{2/3}\frac{\hbar^2}{m}|\psi|^{4/3}\psi.
\end{equation}
In comparison, a self-gravitating BEC with a potential (\ref{mfgp5}) is described by the GP equation
\begin{equation}
\label{analo2}
i\hbar \frac{\partial\psi}{\partial t}=-\frac{\hbar^2}{2m}\Delta\psi+m\Phi\psi+ N\frac{4\pi a\hbar^2}{m}|\psi|^{2}\psi.
\end{equation}
Apart from the difference in the exponent, we see that the  $|\psi|^{4}$ (quartic) self-interaction of bosons ($p\propto \rho^2\propto |\psi|^{4}$) plays a role similar to the Pauli exclusion principle for fermions, equivalent to a $|\psi|^{10/3}$ interaction ($p\propto \rho^{5/3}\propto |\psi|^{10/3}$). Indeed, a quartic self-interaction with $a>0$  is equivalent to an inter-particle repulsion that dominates over uncertainty pressure for $N\gg 1$. Similarly, in the case of fermions, the exclusion pressure dominates over uncertainty pressure for $N\gg 1$. In the Newtonian regime, and in the TF approximation, self-coupled boson stars are equivalent to $n=1$ polytropes and fermion stars (like classical white dwarf stars) to $n=3/2$ polytropes. The analogy between  boson stars with a self-interaction and fermion stars takes even more sense in the relativistic regime \cite{colpi}.  In that case, a self-coupled boson star has an equation of state  $p=\epsilon/3=Kn^{4/3}$ (where $\epsilon$ is the energy density and $n$ is the particle density) corresponding to a $n=3$ polytrope like in the core of neutron stars \cite{ov,mz,htww,chavrelat}.

We note that the potential $h(\rho)\propto \rho^{2/d}$ associated with fermions becomes equivalent to the potential $h(\rho)\propto \rho$ associated with self-coupled bosons when $d=2$. In fact, the dimension $d=2$ is a critical dimension \cite{kolomeisky}. When we consider a gas of repulsive (impenetrable) bosons, the potential $h(\rho)=g\rho$ arising in the GP equation ceases to be valid for $d\le 2$ (in $d=2$ it remains marginally valid with logarithmic corrections). In particular, in $d=1$, it is replaced by $h(\rho)=\pi^2\hbar^2\rho^2/2m^4$ exactly like for spinless ($s=0$) fermions. This is a manifestation of the boson-fermion duality in one dimension \cite{girardeau}.

\section{Exact results}
\label{sec_exact}

\subsection{The energy functional}
\label{sec_ef}

The total energy associated with the GPP system (\ref{mfgp6})-(\ref{mfgp7}) or, equivalently, with the quantum barotropic Euler-Poisson system (\ref{mad3}), (\ref{mad7}) and (\ref{mfgp7})  can be written
\begin{eqnarray}
\label{ef1}
E_{tot}=\Theta_c+\Theta_Q+U+W.
\end{eqnarray}
The first two terms correspond to the total kinetic energy
\begin{eqnarray}
\label{ef2}
\Theta=\frac{N\hbar^2}{2m}\int |\nabla\psi|^2 \, d{\bf r}.
\end{eqnarray}
Using the Madelung transformation, it can be decomposed into the  ``classical'' kinetic energy
\begin{eqnarray}
\label{ef3}
\Theta_c=\int\rho \frac{{\bf u}^2}{2}\, d{\bf r},
\end{eqnarray}
and the ``quantum'' kinetic energy
\begin{equation}
\label{ef4}
\Theta_Q=\frac{1}{m}\int \rho Q\, d{\bf r}.
\end{equation}
Substituting Eq. (\ref{mad5}) in Eq. (\ref{ef4}), we obtain the equivalent expressions
\begin{eqnarray}
\label{ef5}
\Theta_Q&=&-\frac{\hbar^2}{2m^2}\int \sqrt{\rho}\Delta\sqrt{\rho}\, d{\bf r}\nonumber\\
&=&\frac{\hbar^2}{2m^2}\int (\nabla\sqrt{\rho})^2\, d{\bf r}
=\frac{\hbar^2}{8m^2}\int \frac{(\nabla\rho)^2}{\rho}\, d{\bf r}.
\end{eqnarray}
The third term is the internal energy
\begin{eqnarray}
\label{ef6}
U&=&\int\rho\int^{\rho}\frac{p(\rho_1)}{\rho_1^2}\, d\rho_1\, d{\bf r}\nonumber\\
&=&\int \left\lbrack \rho h(\rho)-p(\rho)\right \rbrack\, d{\bf r}=\int H(\rho)\, d{\bf r},
\end{eqnarray}
where we have used Eq. (\ref{mad8}) and integrated by parts to obtain the second equality.
For a polytropic equation of state (\ref{mad10}), it takes the form
\begin{eqnarray}
\label{ef7}
U=\frac{K}{\gamma-1}\int \rho^{\gamma}\, d{\bf r}=\frac{1}{\gamma-1}\int p\, d{\bf r}.
\end{eqnarray}
In particular, for the potential (\ref{mfgp5}), using Eq. (\ref{mad9}), we get
\begin{eqnarray}
\label{ef8}
U=\frac{2\pi a\hbar^2}{m^3}\int \rho^2\, d{\bf r}.
\end{eqnarray}
Finally, the fourth term is the gravitational energy
\begin{eqnarray}
\label{ef9}
W=\frac{1}{2}\int\rho\Phi\, d{\bf r}.
\end{eqnarray}
The total energy per particle can be expressed in terms of the wave function as
\begin{eqnarray}
\label{ef10}
\hat{E}_{tot}=\int \left\lbrack \frac{\hbar^2}{2m}|\nabla\psi|^2+\hat{H}(Nm|\psi|^2)+\frac{1}{2}m\Phi|\psi|^{2}\right\rbrack \, d{\bf r},\nonumber\\
\end{eqnarray}
where $\hat{H}=H/N$. Then, the  GP equation (\ref{mfgp6}) can be written
\begin{eqnarray}
\label{ef11}
i\hbar \frac{\partial \psi}{\partial t}=\frac{\delta \hat{E}_{tot}}{\delta\psi^*}.
\end{eqnarray}
For a power-law potential  $h(\rho)=\lbrack K\gamma/(\gamma-1)\rbrack \rho^{\gamma-1}$, we have $H(\rho)=\lbrack K/(\gamma-1)\rbrack \rho^{\gamma}$ so that $\hat{H}(Nm|\psi|^2)=\kappa |\psi|^{2\gamma}$ with  $\kappa=KN^{\gamma-1}m^{\gamma}/(\gamma-1)$.

It can be shown (see Appendix \ref{sec_conservationE}) that the total energy  $E_{tot}$ is conserved by the quantum barotropic Euler-Poisson system (or by the GPP system).   The mass $M=\int \rho\, d{\bf r}$ is also conserved. Therefore,  a minimum of the energy functional $E_{tot}[\rho,{\bf u}]$ at fixed mass $M$ is a nonlinearly dynamically stable steady state of the quantum barotropic Euler-Poisson system (this follows from general results of dynamical stability \cite{holm}). We are therefore led to considering the minimization problem
\begin{eqnarray}
\label{ef12}
\min_{\rho,{\bf u}} \left\lbrace E_{tot}[\rho,{\bf u}]\quad |\quad M\right\rbrace.
\end{eqnarray}
An extremum of energy at fixed mass is given by the variational
principle $\delta E_{tot}-\alpha\delta M=0$ where $\alpha$ is a
Lagrange multiplier taking into account the mass constraint. Using the
results of Appendix \ref{sec_conservationE}, this gives ${\bf u}={\bf
0}$ and the condition
\begin{eqnarray}
\label{ef14}
m\Phi+m h(\rho)-\frac{\hbar^2}{2m}\frac{\Delta\sqrt{\rho}}{\sqrt{\rho}}=m\alpha.
\end{eqnarray}
This equation is equivalent to the steady state equation (\ref{tigp3}) provided that we make the identification
\begin{eqnarray}
\label{ef15}
\alpha=E/m.
\end{eqnarray}
This shows that the Lagrange multiplier (chemical potential) in the constrained minimization problem (\ref{ef14}) is equal to the eigenenergy $E$ by unit of mass (if we apply the variational principle (\ref{ef12}) at Eq. (\ref{ef10}) we get Eq. (\ref{tigp2})). On the other hand, considering the second order variations of energy, we find that the distribution is dynamically stable iff
\begin{eqnarray}
\label{ef16}
\delta^2 E_{tot}\equiv \frac{1}{2}\int h'(\rho)(\delta\rho)^2\, d{\bf r}+\frac{1}{2}\int \delta\rho\delta\Phi\, d{\bf r}\nonumber\\
+\frac{\hbar^2}{8m^2}\int  \left \lbrack \nabla \left (\frac{\delta\rho}{\sqrt{\rho}}\right )\right\rbrack^2\, d{\bf r}+\frac{\hbar^2}{8m^2}\int \frac{\Delta\sqrt{\rho}}{\rho^{3/2}}(\delta\rho)^2\, d{\bf r}>0,\nonumber\\
\end{eqnarray}
for all perturbations that conserve mass: $\int \delta\rho\, d{\bf r}=0$.

{\it Remark 1:} If we plot $\alpha=E/m=\partial E_{tot}/\partial M$ (conjugate quantity) as a function of $M$ (conserved quantity), we can determine the stability of the system by a direct application of the Poincar\'e theory of linear series of equilibria (see, e.g. \cite{katz,ijmpb} for details). According to the Poincar\'e theorem, a change of stability can only occur at a turning point of mass  or at a bifurcation point. Therefore, if we know a limit in which the configuration is stable, then we can use the Poincar\'e theorem to deduce the stability of the whole series of equilibria. We shall use this method in Paper II.

{\it Remark 2:} In the TF approximation, the energy functional (\ref{ef1}) reduces to the standard Chandrasekhar energy functional. It is well-known that a polytrope with index $\gamma>4/3$, including the polytrope $\gamma=2$ ($n=1$) corresponding to Eq. (\ref{mad9}), is a minimum of energy at fixed mass. Therefore, it is nonlinearly dynamically stable with respect to the barotropic Euler-Poisson system. Its linear dynamical stability can also be settled by using the Eddington \cite{eddingtonpuls} equation of pulsation or the Ledoux \cite{ledoux} criterion (see \cite{bt} and Appendix B of \cite{prd}).

\subsection{The virial theorem}
\label{sec_virial}

From the quantum barotropic Euler-Poisson system  (\ref{mad3}), (\ref{mad7}) and (\ref{mfgp7}), we can derive the general time-dependent virial theorem (see Appendix \ref{sec_virialannexe}):
\begin{equation}
\label{v1}
\frac{1}{2}\ddot I=2(\Theta_c+\Theta_Q)+3\int p\, d{\bf r}+W,
\end{equation}
where
\begin{equation}
\label{v2}
I=\int \rho r^2\, d{\bf r},
\end{equation}
is the moment of inertia. For a polytropic equation of state (\ref{mad10}), we have the identity  $\int p\, d{\bf r}=(\gamma-1)U$. More specifically, for the potential (\ref{mfgp5}) leading to Eq. (\ref{mad9}), we get $\int p\, d{\bf r}=U$. In that case, the time-dependent virial theorem can be rewritten
\begin{equation}
\label{v3}
\frac{1}{2}\ddot I=2(\Theta_c+\Theta_Q)+3U+W.
\end{equation}
At equilibrium ($\ddot I=\Theta_c=0$), we obtain the time-independent virial theorem
\begin{equation}
\label{v4}
2\Theta_Q+3U+W=0.
\end{equation}
On the other hand, the energy functional (\ref{ef1}) reduces to
\begin{eqnarray}
\label{v5}
E_{tot}=\Theta_Q+U+W.
\end{eqnarray}
Finally, multiplying the steady state equation (\ref{tigp4})  by $\rho$ and integrating over the configuration, we obtain the general identity
\begin{equation}
\label{v6}
\Theta_Q+\int \rho h\, d{\bf r}+2W=NE.
\end{equation}
For a polytropic equation of state (\ref{mad10}), we find that $\int \rho h\, d{\bf r}=\gamma U$. More specifically, for the potential (\ref{mfgp5}) leading to Eq. (\ref{mad9}), we get $\int \rho h\, d{\bf r}=2U$. In that case, Eq. (\ref{v6}) can be rewritten
\begin{equation}
\label{v7}
\Theta_Q+2 U+2W=NE.
\end{equation}

In the non-interacting case ($a=0$), the internal energy vanishes: $U=0$. The three independent equations (\ref{v4}), (\ref{v5}) and (\ref{v7}) reduce to $2\Theta_Q+W=0$, $E_{tot}=\Theta_Q+W$ and $\Theta_Q+2W=NE$.
From these equations, we obtain the relation
\begin{equation}
\label{v11}
E_{tot}=\frac{1}{3}NE.
\end{equation}
This relation shows that the total energy $E_{tot}$ is not equal to $NE$, as we could naively believe \cite{rb}. The $1/3$ factor was previously obtained  by Membrado {\it et al.} \cite{membrado} from a different argument.
We can check that the above relations are satisfied by the different components (\ref{nic3})-(\ref{nic6}) of the  energy.

In the TF approximation, the quantum energy is neglected: $\Theta_Q=0$. The three independent equations (\ref{v4}), (\ref{v5}) and (\ref{v7}) reduce to $3U+W=0$, $E_{tot}=U+W$ and $2 U+2W=NE$. From these equations, we obtain the relation
\begin{equation}
\label{v15}
E_{tot}=\frac{1}{2}NE.
\end{equation}
We can check that the above relations are satisfied by the different components (\ref{tf8}), (\ref{tf10}), (\ref{tf11}) and (\ref{tf12}) of the  energy.

Finally, in the non-gravitational limit, the potential energy is neglected: $W=0$. The three independent equations (\ref{v4}), (\ref{v5}) and (\ref{v7}) reduce to $2\Theta_Q+3U=0$, $E_{tot}=\Theta_Q+U$ and $\Theta_Q+2 U=NE$. From these equations, we obtain the relation
\begin{equation}
\label{v16}
E_{tot}=-NE.
\end{equation}

\section{The Gaussian Ansatz}
\label{sec_ansatz}

To obtain the density profile of a self-gravitating BEC and the mass-radius relation, we need to solve the differential equation (\ref{he3}) expressing the condition of hydrostatic equilibrium. This will be done numerically in Paper II. However, it can also be useful to obtain approximate analytical results. In that respect, we shall follow an approach similar to the one developed by Stoner (1929,1930) \cite{stonerclassic,stoner}, Nauenberg (1972) \cite{nauenberg} and Chavanis (2007) \cite{prd} in the case of classical and relativistic white dwarf stars. The idea is to prescribe an approximate density profile (characterized by its mass $M$ and radius $R$), compute the total energy $E_{tot}(R)$ and minimize it with respect to $R$, for a given mass $M$,  in order to obtain the equilibrium radius $R(M)$. This method has provided very good approximations of the mass-radius relation of white dwarf stars. We shall see that it also provides a good approximation of the mass-radius relation of self-gravitating BECs. In order to evaluate the quantum kinetic energy which involves density gradients, we shall make a Gaussian ansatz. The Gaussian ansatz is particularly accurate for small or moderate values of $\chi$ for which the density profile extends to infinity (e.g., in the non-interacting case $\chi=0$). By contrast, in the case $\chi\gg 1$, this ansatz is poor because the density profile approaches that of a $n=1$ polytrope which has a compact support (this is the exact solution in the TF limit $\chi\rightarrow +\infty$). However, for the sake of illustration, we shall use the Gaussian ansatz for all configurations.

\subsection{The energy functional}
\label{sec_efg}

We shall calculate the energy functional (\ref{ef1}) by making a Gaussian ansatz for the density profile
\begin{eqnarray}
\label{efg1}
\rho(r)=M\left (\frac{1}{\pi R^2}\right )^{3/2}e^{-\frac{r^2}{R^2}}.
\end{eqnarray}
The central density is $\rho(0)={M}/({\pi^{3/2} R^3})$ and the corresponding rotation curve is
\begin{eqnarray}
\label{efg2}
v_c(r)=\left (\frac{GM}{R}\right )^{1/2}\left\lbrack \frac{R}{r}\erf\left (\frac{r}{R}\right )-\frac{2}{\sqrt{\pi}}e^{-\left (\frac{r}{R}\right )^2}\right\rbrack^{1/2}.
\end{eqnarray}
For comparison with the results of Paper II, it is convenient to introduce the radius containing $99\%$ of the total mass. For the Gaussian density profile (\ref{efg1}), we find that $R_{99}= 2.38167R$. On the other hand, using Eq. (\ref{efg1}), the quantum kinetic energy, the internal energy and the potential energy are given by
\begin{eqnarray}
\label{efg3}
\Theta_Q=\sigma \frac{\hbar^2M}{m^2R^2},\quad U={\zeta}\frac{2\pi a \hbar^2 M^2}{m^3 R^3},\quad W=-\nu \frac{GM^2}{R},\nonumber\\
\end{eqnarray}
with the coefficients
\begin{eqnarray}
\label{efg4}
\sigma=\frac{3}{4},\qquad \zeta=\frac{1}{(2\pi)^{{3}/{2}}},\qquad \nu=\frac{1}{\sqrt{2\pi}}.
\end{eqnarray}
To compute the potential energy, we have used the formula $W=-4\pi G\int_0^{+\infty} \rho(r)M(r)r\, dr$
valid for a spherically symmetric distribution of matter, and integrated by parts. The moment of inertia is given by
\begin{eqnarray}
\label{efg5}
I=\alpha MR^2, \quad {\rm with}\quad \alpha=\frac{3}{2}.
\end{eqnarray}
Finally, using the velocity profile  (\ref{velf5}) of Appendix \ref{sec_velf}, we find that the classical kinetic energy is given by
\begin{eqnarray}
\label{efg6}
\Theta_c=\frac{1}{2}\alpha M\left (\frac{dR}{dt}\right )^2.
\end{eqnarray}

\begin{figure}[!h]
\begin{center}
\includegraphics[clip,scale=0.3]{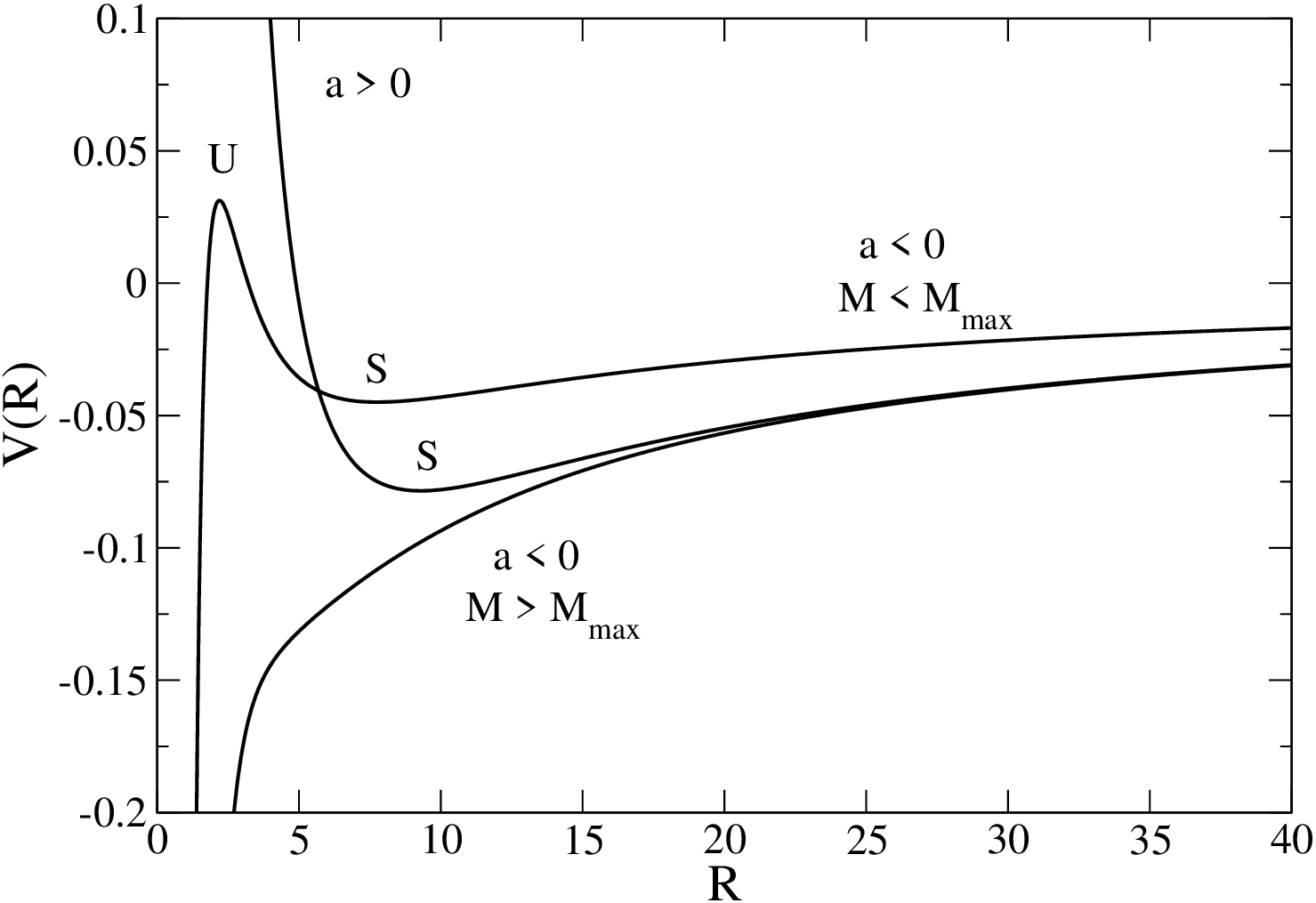}
\caption{Potential $V(R)$ for different values of $M$. The radius and the mass are normalized by $R_a$ and $M_a$ (with $a$ replaced by $|a|$) and the potential by $E_a=GM_a^2/R_a=\hbar (Gm)^{1/2}/|a|^{3/2}$. Then, $V(R)=\sigma M/R^2\pm 2\pi\zeta M^2/R^3-\nu M^2/R$ (with $+$ for $a>0$ and $-$ for $a<0$). For $a>0$, we have taken $M=1.2$ and for $a<0$ we have taken $M=0.9<M_{max}$ and $M=1.2>M_{max}$ where $M_{max}= 1.085$ (see below).}
\label{potentiel}
\end{center}
\end{figure}

From these expressions, the energy functional (\ref{ef1}) can be rewritten as a function of $R$ and $\dot R$  (for a fixed mass $M$) as
\begin{eqnarray}
\label{efg7}
E_{tot}=\frac{1}{2}\alpha M\left (\frac{dR}{dt}\right )^2+\sigma \frac{\hbar^2M}{m^2R^2}+{\zeta}\frac{2\pi a \hbar^2 M^{2}}{m^3R^{3}}-\nu \frac{GM^2}{R}.\nonumber\\
\end{eqnarray}
This can be interpreted as the total energy
\begin{eqnarray}
\label{efg8}
E_{tot}=\frac{1}{2}\alpha M\left (\frac{dR}{dt}\right )^2+V(R),
\end{eqnarray}
of a fictive   particle with mass $\alpha M$ and position $R$  moving in a potential
\begin{eqnarray}
\label{efg9}
V(R)=\sigma \frac{\hbar^2M}{m^2R^2}+{\zeta}\frac{2\pi a \hbar^2 M^{2}}{m^3R^{3}}-\nu \frac{GM^2}{R}.
\end{eqnarray}
We shall come back on this mechanical analogy in Sec. \ref{sec_vg}. The potential $V(R)$ is plotted in Fig. \ref{potentiel} for different values of $M$ (with $a>0$ and $a<0$). These different cases are studied in the sequel.

\subsection{The mass-radius relation}
\label{sec_mr}

A minimum of the energy functional $E_{tot}[\rho,{\bf u}]$, given by Eq. (\ref{ef1}), at fixed mass $M$ determines a stable steady state of the quantum barotropic Euler-Poisson system (equivalent to the Gross-Pitaevskii-Poisson system). In the Gaussian approximation, we are led to determining the minimum of the function $E_{tot}(R,\dot R)$, given by Eq. (\ref{efg8}), at fixed mass $M$. Clearly, we must have $\dot R=0$ so that a minimum of energy at fixed mass is a steady state. Then, we must determine the minimum of the potential energy $V(R)$. Taking the derivative of ${V}(R)$ with respect to $R$, we obtain
\begin{eqnarray}
\label{mr1}
{V}'(R)=
-2\sigma\frac{\hbar^2 M}{m^2 R^3}-3\zeta \frac{2\pi a \hbar^2 M^{2}}{m^3R^{4}}+\nu \frac{G M^2}{R^2}.
\end{eqnarray}
The  critical points of the potential energy $V(R)$, satisfying   $d{V}/dR=0$, are solution of the equation
\begin{eqnarray}
\label{mr2}
M=\frac{2\sigma}{\nu}\frac{\frac{\hbar^2}{Gm^2 R}}{1-\frac{6\pi\zeta a\hbar^2}{\nu Gm^3R^2}},
\end{eqnarray}
The radius can be expressed as a function of the mass according to
\begin{eqnarray}
\label{fa5inv}
R=\frac{\sigma}{\nu}\frac{\hbar^2}{GMm^2}\left (1\pm\sqrt{1+\frac{6\pi\zeta \nu}{\sigma^2}\frac{GmM^2a}{\hbar^2}}\right ),
\end{eqnarray}
with $+$ when $a\ge 0$ and $\pm$ when $a<0$. This equation provides an analytical approximate expression of the mass-radius relation of a self-gravitating BEC with short-range interactions. It can also be obtained from the equilibrium virial theorem (\ref{v4}) by making the Gaussian ansatz (see Sec. \ref{sec_vg}).

Let us first consider asymptotic limits of this relation. (i) In the non-interacting case ($a=0$), we obtain
\begin{eqnarray}
\label{mr3}
R=\frac{2\sigma}{\nu}\frac{\hbar^2}{GMm^2}.
\end{eqnarray}
The radius $R_{99}$ containing $99\%$ of the mass is $R_{99}=8.955{\hbar^2}/{GMm^2}$. This can be compared with the exact result (\ref{nic2}) giving $R_{99}^{exact}=9.9{\hbar^2}/{GMm^2}$. The agreement is fairly good. (ii) In the TF approximation (when $a>0$), we get
\begin{eqnarray}
\label{mr4}
R=\left (\frac{6\pi\zeta}{\nu}\right )^{1/2}\left (\frac{a\hbar^2}{Gm^3}\right )^{1/2}.
\end{eqnarray}
The radius is independent on mass. The radius $R_{99}$ containing $99\%$ of the mass is given by $R_{99}=4.125 ({a\hbar^2}/{Gm^3})^{1/2}$. This can be compared with the exact result (\ref{tf5}) giving $R_{99}^{exact}=2.998 ({a\hbar^2}/{Gm^3})^{1/2}$. The agreement is less good than in the non-interacting case. The reason is related to the fact that the distribution (\ref{tf4}) has a compact support so that it is quite different from a Gaussian. The radius defined by Eq. (\ref{mr4}) represents the minimum radius $R_{min}$ of the self-gravitating BEC for a given value of the scattering length $a>0$. For $R\rightarrow R_{min}$, the mass tends to $+\infty$. According to Eq. (\ref{mr2}), it diverges like
\begin{eqnarray}
\label{mr5}
M\sim \left (\frac{\sigma^2}{6\pi\zeta\nu}\right )^{1/2}\frac{\hbar}{\sqrt{G m a}}\frac{1}{\frac{R}{R_{min}}-1}.
\end{eqnarray}
(iii) In the non-gravitational limit (when $a<0$), we get
\begin{eqnarray}
\label{mr6}
R=\frac{3\pi \zeta}{\sigma}\frac{M |a|}{m}.
\end{eqnarray}
The radius $R_{99}$ containing $99\%$ of the mass is given by $R_{99}=1.900 M |a|/m$. However, we shall see that these configurations are always unstable.

A critical point of ${V}(R)$ is an energy minimum iff ${V}''(R)>0$. Computing the second derivative of ${V}(R)$, we obtain
\begin{eqnarray}
\label{mr7}
{V}''(R)=6\sigma\frac{\hbar^2 M}{m^2 R^4}+12\zeta \frac{2\pi a \hbar^2 M^{2}}{m^3R^{5}}-2\nu \frac{G M^2}{R^3}.
\end{eqnarray}
Using the mass-radius relation (\ref{mr2}), the foregoing equation can be rewritten
\begin{eqnarray}
\label{mr8}
{V}''(R)=\frac{\nu GM^2}{R^3}\left (1+\frac{6\pi \zeta a \hbar^2}{\nu G m^3R^2}\right ).
\end{eqnarray}

\begin{figure}[!h]
\begin{center}
\includegraphics[clip,scale=0.3]{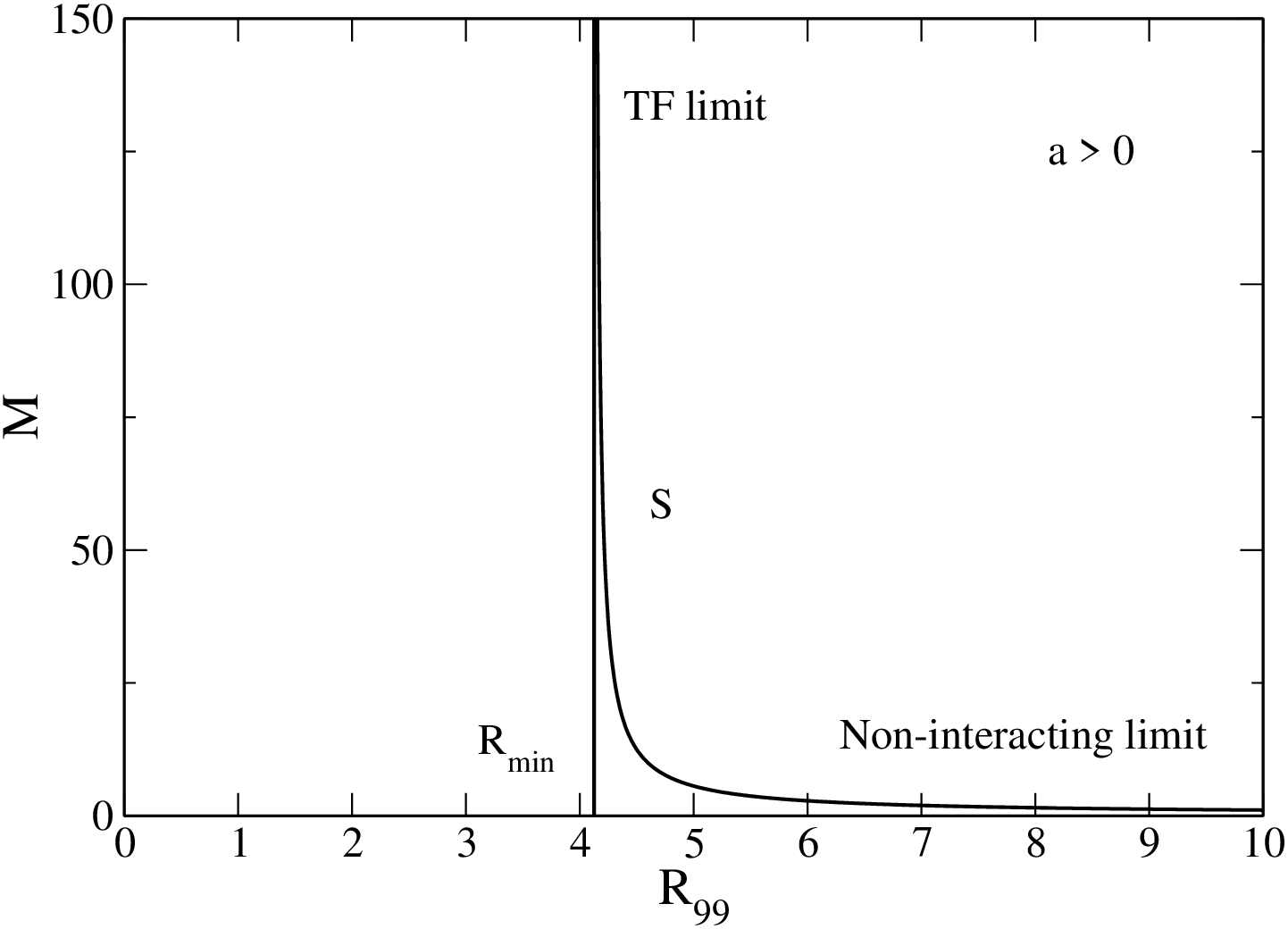}
\caption{$M$ as a function of $R_{99}$ for given $a>0$. The mass is normalized by $M_a$ and the radius by $R_a$. Thus,  $M=2\sigma R/(\nu R^2-6\pi\zeta)$ with $R_{99}=2.38167R$. The radius is given as a function of the mass by $R=(\sigma/\nu M)(1+\sqrt{1+6\pi\zeta\nu M^2/\sigma^2})$. In the non-interacting limit $M\rightarrow 0$, we get $M\sim 2\sigma/\nu R$ i.e. $M\sim 8.955/R_{99}$ and in the TF limit $M\rightarrow +\infty$, we get $R\rightarrow (6\pi\zeta/\nu)^{1/2}$ i.e. $R_{99}\rightarrow 4.125$. All the configurations are stable.}
\label{M-R-a-pos}
\end{center}
\end{figure}

\begin{figure}[!h]
\begin{center}
\includegraphics[clip,scale=0.3]{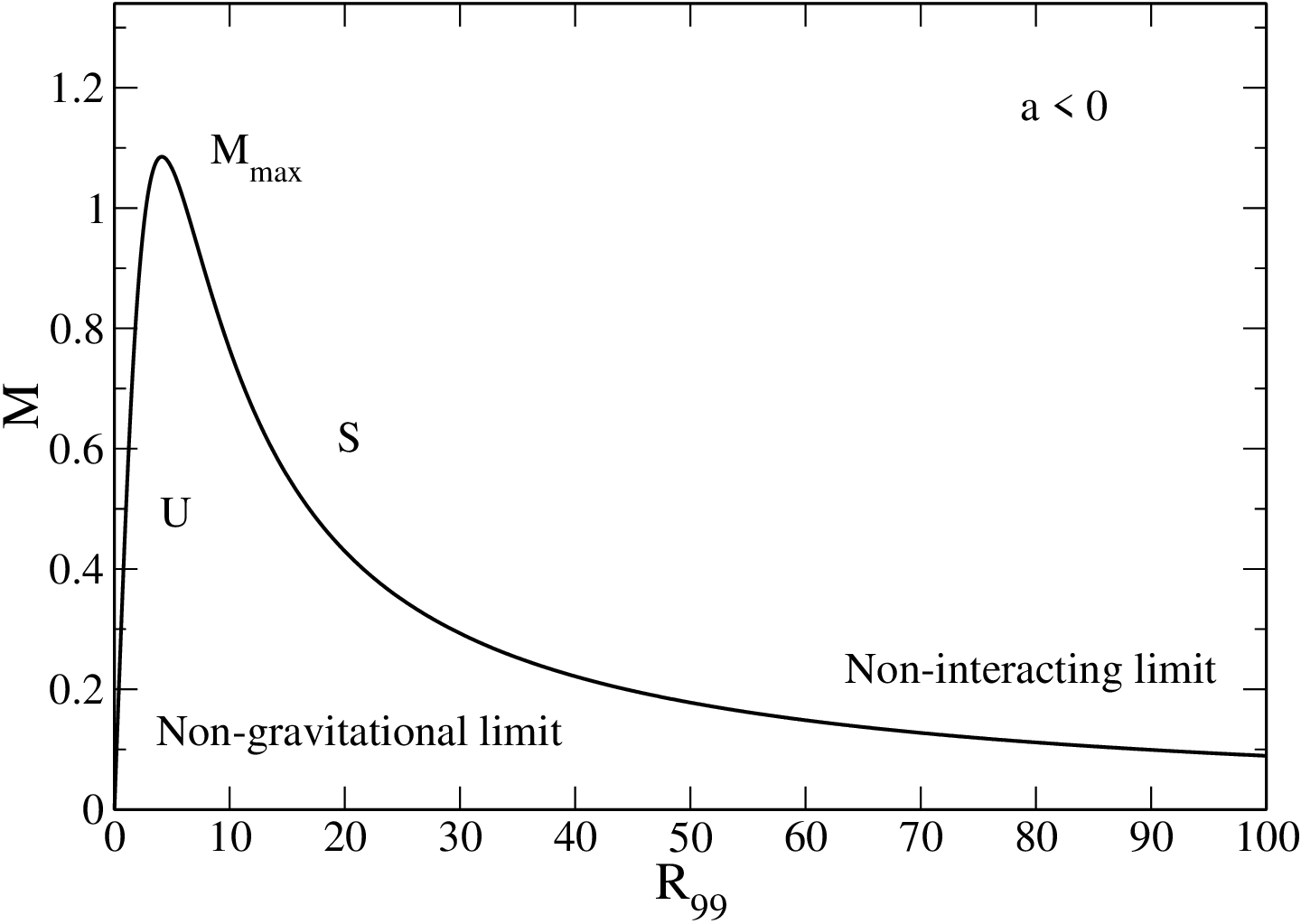}
\caption{${M}$ as a function of ${R}_{99}$ for given $a<0$. The mass is normalized by $M_a$  and the radius by $R_a$ (where $a$ is replaced by $|a|$). Thus, $M=2\sigma R/(\nu R^2+6\pi\zeta)$ with $R_{99}=2.38167R$. The radius is given as a function of the mass by $R=(\sigma/\nu M)(1\pm\sqrt{1-6\pi\zeta\nu M^2/\sigma^2})$. In the non-interacting limit  $R\rightarrow +\infty$, we get $M\sim 2\sigma/\nu R$ i.e. $M\sim 8.955/R_{99}$ and in the non-gravitational limit $R\rightarrow 0$ (unstable), we get $M\sim \sigma R/(3\pi\zeta)$ i.e. $M\sim 0.5262 R_{99}$. There exists a maximum mass $M_{max}=\sigma/\sqrt{6\pi\zeta\nu}= 1.085$ corresponding to a radius $R_*=(6\pi\zeta/\nu)^{1/2}$ i.e. $R_{99}^*= 4.125$.  The configurations are stable for $R>R_*$ and unstable for $R<R_*$.}
\label{M-R-neg}
\end{center}
\end{figure}

Let us first consider the case of repulsive short-range interactions ($a>0$). The mass-radius relation is represented in Fig. \ref{M-R-a-pos}. There exists one, and only one, solution for each value of the mass and it is stable since, according to Eq. (\ref{mr8}), it is a minimum of energy ($V''(R)>0$); see Fig. \ref{potentiel}. The radius is a decreasing function of the mass. The non-interacting limit corresponds to $M\ll M_a$, i.e. $R\gg  R_a\sim R_{min}$. For $M\rightarrow 0$, the radius $R\rightarrow +\infty$ with the scaling (\ref{mr3}). The TF limit corresponds to $M\gg M_a$, i.e. $R\sim R_a\sim R_{min}$. For $M\rightarrow +\infty$, the radius $R$ tends to the minimum value $R_{min}$ given by Eq. (\ref{mr4})  with the scaling (\ref{mr5}). In the non-interacting case $a=0$, the mass-radius relation is given by Eq. (\ref{mr3}) which determines a stable equilibrium state for any mass $M$.

Let us now consider the case of attractive short-range interactions ($a<0$). The mass-radius relation is represented in Fig. \ref{M-R-neg}.  There exists a maximum mass
\begin{eqnarray}
\label{mr9}
M_{max}=\left (\frac{\sigma^2}{6\pi\zeta\nu}\right )^{1/2}\frac{\hbar}{\sqrt{Gm|a|}},
\end{eqnarray}
corresponding to the radius
\begin{eqnarray}
\label{mr10}
R_*=\left (\frac{6\pi\zeta}{\nu}\right )^{1/2}\left (\frac{|a|\hbar^2}{Gm^3}\right )^{1/2}.
\end{eqnarray}
It may be noted that, contrary to the Chandrasekhar mass or to the Kaup mass, this maximum mass is a purely Newtonian result. For that reason, it is generally very small (see Appendix \ref{sec_att}). The approximate values $M_{max}=1.085\hbar/\sqrt{Gm|a|}$ and $R_{99}^*=4.125(|a|\hbar^2/Gm^3)^{1/2}$ obtained with the Gaussian ansatz are in fairly good agreement with the exact results $M_{max}^{exact}=1.012\hbar/\sqrt{Gm|a|}$ and $(R^*_{99})^{exact}=5.5(|a|\hbar^2/Gm^3)^{1/2}$ obtained numerically in Paper II. For $M>M_{max}$, there is no solution (no critical point of energy) and the system undergoes gravitational collapse to a black hole. For $M<M_{max}$, there exists two solutions with the same mass. However, according to Eq. (\ref{mr8}), only the solution with the largest radius $R>R_*$ is stable (local minimum of energy $V''(R)>0$). The other solution is an unstable maximum of energy ($V''(R)<0$); see Fig. \ref{potentiel}. We can check that the change of stability ($V''(R)=0$) occurs at the turning point of mass in the series of equilibria ($M'(R)=0$), in agreement with the Poincar\'e theorem. In the stable region, the radius is a decreasing function of the mass.  The non-interacting limit corresponds to $M\ll M_a\sim M_{max}$ and $R\gg R_a\sim R_{*}$.  For $M\rightarrow 0$, the radius $R\rightarrow +\infty$ with the scaling (\ref{mr3}). For $M\rightarrow M_{max}$, the radius $R$ tends to the minimum value $R_*$. The non-gravitational limit corresponds to $M\rightarrow 0$ and $R\rightarrow 0$ but these configurations are inaccessible since they are dynamically unstable (energy maxima) \footnote{In the absence of gravitational forces, BECs with $a<0$ are always unstable.}. If the system is initially placed on the unstable branch, it is expected to undergo gravitational collapse ($R(t)\rightarrow 0$) or to evaporate ($R(t)\rightarrow +\infty$), see Fig. \ref{potentiel}. It may also relax towards the stable equilibrium state with a larger radius ($R(t)\rightarrow R_S$) provided that it is able to dissipate energy, e.g. by radiation.

\subsection{The virial theorem}
\label{sec_vg}

Using the Gaussian ansatz, the time-dependent virial theorem (\ref{v3}) can be written
\begin{eqnarray}
\label{vg1}
\frac{1}{2}\alpha M\frac{d^2R^2}{dt^2}
=\alpha M\left (\frac{dR}{dt}\right )^2+2\sigma\frac{\hbar^2 M}{m^2 R^2}\nonumber\\
+3\zeta \frac{2\pi a \hbar^2 M^{2}}{m^3 R^{3}}-\nu\frac{GM^2}{R}.
\end{eqnarray}
Since
\begin{eqnarray}
\label{vg2}
\frac{d^2R^2}{dt^2}=2R\frac{d^2R}{dt^2}+2\left (\frac{dR}{dt}\right )^2,
\end{eqnarray}
we note the nice cancelation of terms in Eq. (\ref{vg1}) leading to the final equation
\begin{eqnarray}
\label{vg3}
\alpha M\frac{d^2R}{dt^2}
=2\sigma\frac{\hbar^2 M}{m^2 R^3}+3\zeta \frac{2\pi a\hbar^2M^2}{m^3R^{4}}-\nu \frac{G M^2}{R^2}.
\end{eqnarray}
The equilibrium virial theorem ($d^2R/dt^2=0$) returns the mass-radius relation (\ref{mr2}) obtained from the condition $d{V}/dR=0$. In fact, the time-dependent virial theorem (\ref{vg3}) can be written
\begin{eqnarray}
\label{vg4}
\alpha M\frac{d^2R}{dt^2}=-\frac{d{V}}{dR}.
\end{eqnarray}
This equation describes the motion of a fictive particle with mass $\alpha M$ and position $R$ in a potential $V(R)$. Therefore, the total energy $E_{tot}=\Theta_c+V$ given by Eq. (\ref{efg8}) is conserved
\begin{eqnarray}
\label{vg5}
\frac{dE_{tot}}{dt}=\frac{d}{dt}(\Theta_c+V)=0.
\end{eqnarray}
In this mechanical analogy, a stable equilibrium state corresponds to a {\it minimum} of  ${V}(R)$ as we have previously indicated. Alternatively, Eq. (\ref{vg4}) can be viewed as the Hamilton equation of motion of the fictive particle associated with the Hamiltonian (\ref{efg8}).

\subsection{The pulsation equation}
\label{sec_pulse}

To study the linear dynamical stability of a steady state of Eq. (\ref{vg4}), we make a small perturbation around that state and write $R(t)=R+\epsilon(t)$ where $R$ is the equilibrium radius and $\epsilon(t)\ll R$ is the perturbation. Using $V'(R)=0$ and keeping only terms that are linear in $\epsilon$, we obtain the equation
\begin{eqnarray}
\label{pulse1}
\frac{d^2\epsilon}{dt^2}+\omega^2\epsilon=0,
\end{eqnarray}
where $\omega$ is the complex pulsation given by
\begin{eqnarray}
\label{pulse2}
\omega^2=\frac{1}{\alpha M}{V}''(R).
\end{eqnarray}
According to this equation, a steady state is linearly stable iff $\omega^2>0$ that is to say iff it is a (local) minimum of energy $V(R)$. In that case, the system oscillates about its equilibrium value with a pulsation $\omega$. Otherwise, the perturbation grows exponentially rapidly with a growth rate $\lambda=\sqrt{-\omega^2}$. Using  Eq. (\ref{mr7}) for $V''(R)$ and comparing with  Eq. (\ref{efg3}), we find that
\begin{eqnarray}
\label{pulse3}
\omega^2=\frac{6\Theta_Q+12U+2W}{I}.
\end{eqnarray}
In the non-interacting case ($U=0$), using the virial theorem (\ref{v4}), we get $\omega^2=-{W}/{I}$. In the TF approximation ($\Theta_Q=0$), using the virial theorem (\ref{v4}), we obtain $\omega^2=-{2W}/{I}$.
This expression coincides with the Ledoux formula for a polytrope of index $\gamma=2$ \cite{ledoux}. In paper III, we show that this result can be generalized to an arbitrary polytropic index (see also Ref. \cite{prd}).

Using Eqs. (\ref{pulse2}), (\ref{mr8}) and (\ref{mr2}), the pulsation can be written in terms of the radius as
\begin{eqnarray}
\label{pulse6}
\omega^2=\frac{2\sigma}{\alpha}\frac{\hbar^2}{m^2R^4}\frac{1+\frac{6\pi \zeta a \hbar^2}{\nu Gm^3R^2}}{1-\frac{6\pi \zeta a \hbar^2}{\nu Gm^3R^2}}.
\end{eqnarray}

\begin{figure}[!h]
\begin{center}
\includegraphics[clip,scale=0.3]{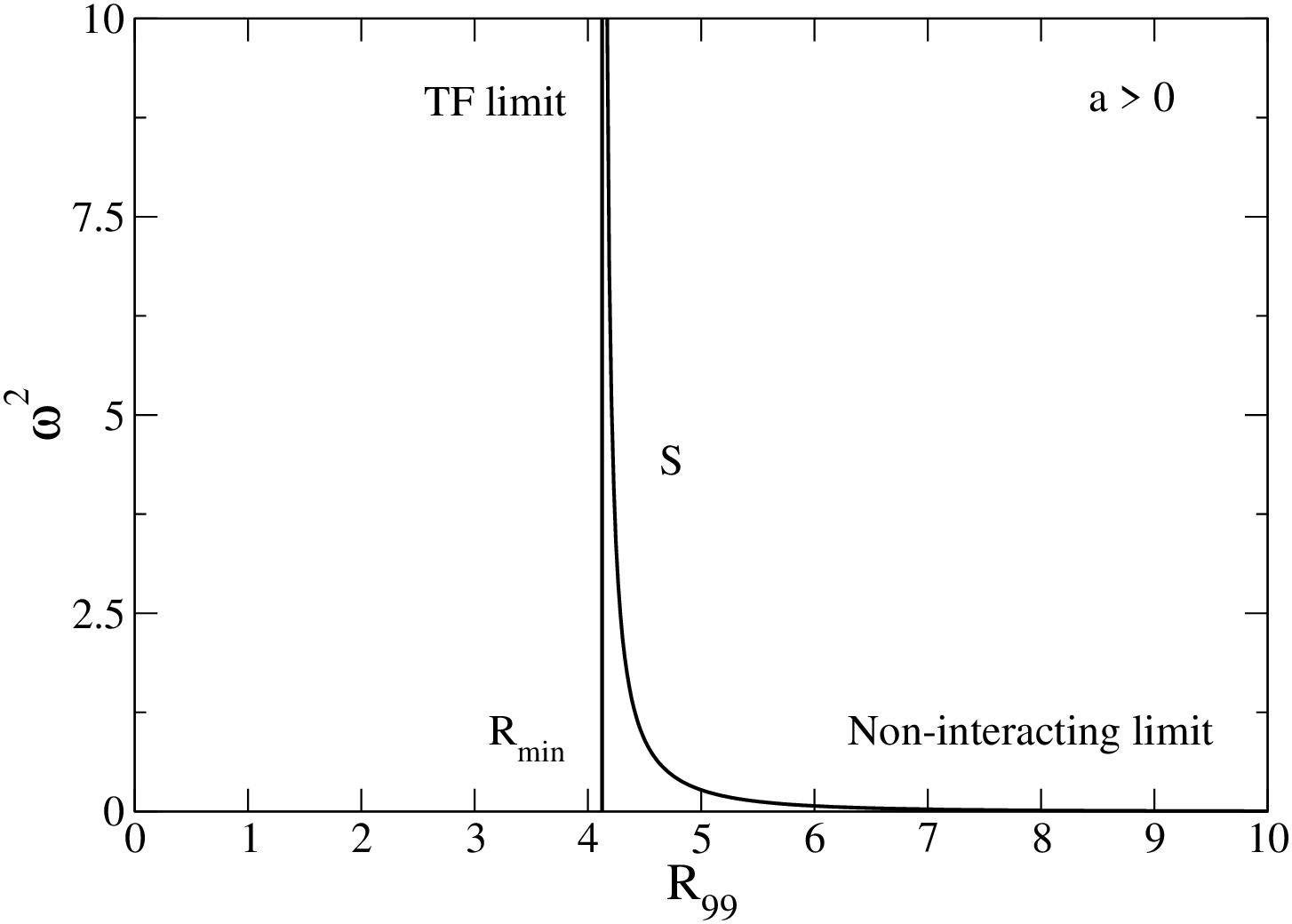}
\caption{$\omega^2$ as a function of $R_{99}$ for given $a>0$. The pulsation is normalized by $\omega_a=(GM_a/R_a^3)^{1/2}=Gm^2/a\hbar$ and the radius by $R_a$. Thus, $\omega^2=(2\sigma/\alpha R^4)(1+6\pi\zeta/\nu R^2)/(1-6\pi\zeta/\nu R^2)$ with $R_{99}=2.38167R$. In the non-interacting limit $R\rightarrow +\infty$, we get $\omega^2\sim 2\sigma/\alpha R^4$ i.e. $\omega^2\sim 32.18/R_{99}^4$ and in the TF limit $R\rightarrow R_{min}$, we get $\omega^2\sim (\sigma\nu^2/18\alpha\pi^2\zeta^2)/(R/R_{min}-1)\rightarrow +\infty$.}
\label{pulsation-aFIXEpos}
\end{center}
\end{figure}

\begin{figure}[!h]
\begin{center}
\includegraphics[clip,scale=0.3]{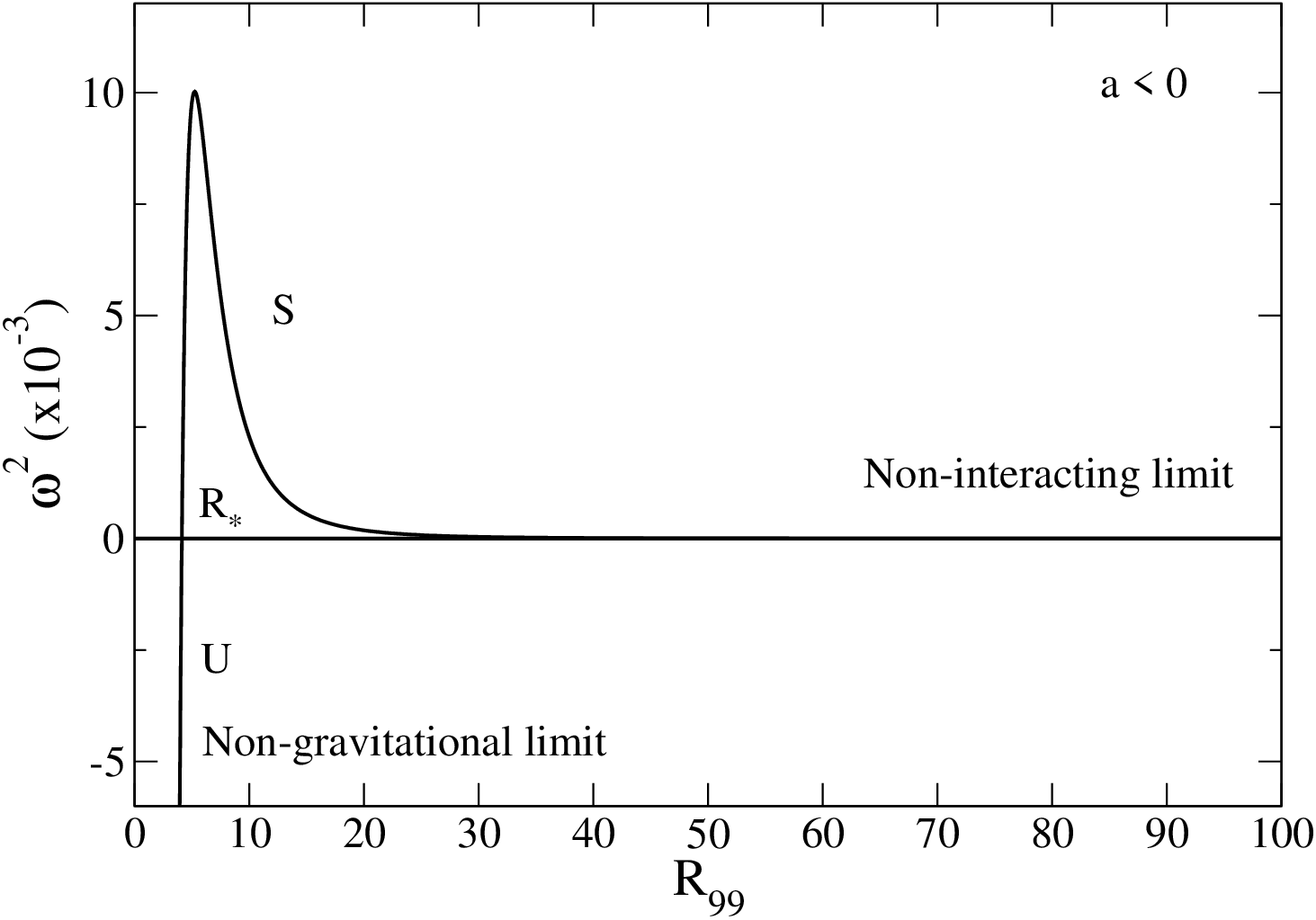}
\caption{$\omega^2$ as a function of $R_{99}$ for given $a<0$. The pulsation is normalized by $\omega_a=(GM_a/R_a^3)^{1/2}=Gm^2/|a|\hbar$ and the radius by $R_a$ (with $a$ replaced by $|a|$). Thus, $\omega^2=(2\sigma/\alpha R^4)(1-6\pi\zeta/\nu R^2)/(1+6\pi\zeta/\nu R^2)$ with $R_{99}=2.38167R$. In the non-interacting limit $R\rightarrow +\infty$, we get  $\omega^2\sim 2\sigma/\alpha R^4$ i.e. $\omega^2\sim 32.18/R_{99}^4$ and in the non-gravitational limit $R\rightarrow 0$ (unstable), we get  $\omega^2\sim -2\sigma/\alpha R^4$ i.e. $\omega^2\sim -32.18/R_{99}^4$. The pulsation is maximum for $R_{99}= 5.247$ with the value $\omega^2= 10.02\, 10^{-3}$.}
\label{pulsation-aFIXEneg}
\end{center}
\end{figure}

\begin{figure}[!h]
\begin{center}
\includegraphics[clip,scale=0.3]{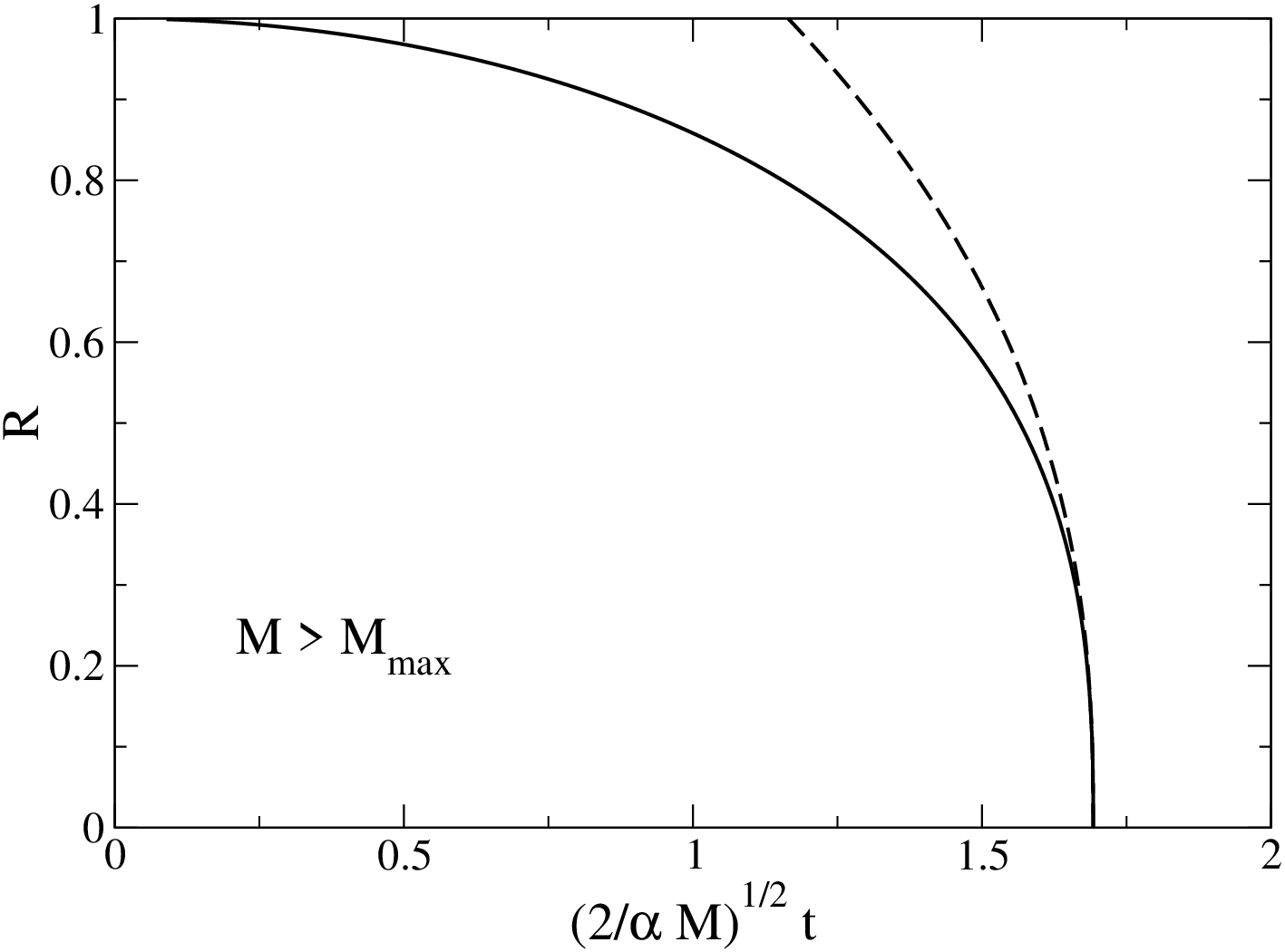}
\caption{Collapse of the BEC for $M>M_{max}$ in the case of an attractive self-interaction ($a<0$).   We have represented the BEC radius $R(t)$ as a function of time, by solving Eq. (\ref{vg3}) obtained with the Gaussian ansatz. The radius is normalized by $R_a$, the mass by $M_a$ and the time by $t_a=|a|\hbar/Gm^2$. The equation of motion can then be rewritten $\alpha M \ddot R=-V'(R)$ with $V(R)=\sigma M/R^2-2\pi\zeta M^2/R^3-\nu M^2/R$. The solution is $(2/\alpha M)^{1/2}t=\int_{R(t)}^{R_0}dR/\sqrt{V(R_0)-V(R)}$ (solid line). The collapse leads to a finite time singularity. The collapse time $t_{coll}$ is obtained from the foregoing expression by using $R(t_{coll})=0$. The solution can then be rewritten $(2/\alpha M)^{1/2}(t_{coll}-t)=\int_{0}^{R(t)}dR/\sqrt{V(R_0)-V(R)}$. For $t\rightarrow t_{coll}$, the radius behaves like $R(t)\simeq (25\pi\zeta M/\alpha)^{1/5}(t_{coll}-t)^{2/5}$ (dashed line). We have taken $M=1.2$, $R_0=1$ yielding $(2/\alpha M)^{1/2}t_{coll}=1.692$.}
\label{collapseBEC}
\end{center}
\end{figure}

For $a>0$, the pulsation-radius relation is plotted in
Fig. \ref{pulsation-aFIXEpos}. We note that $\omega^2>0$ for any $R$
so that the configurations are stable. When slightly perturbed about
its equilibrium state, the system oscillates with a pulsation
$\omega$. For $a<0$, the pulsation-radius relation is plotted in
Fig. \ref{pulsation-aFIXEneg}. We find that $\omega^2>0$ for $R>R_*$
(stable) while $\omega^2<0$ for $R<R_*$ (unstable). In the latter
case, the perturbation grows with a growth rate
$\lambda=\sqrt{-\omega^2}$.

Let us derive asymptotic expressions of the pulsation. (i) In the non-interacting case,  we obtain
\begin{eqnarray}
\label{pulse7}
\omega^2=\frac{\nu}{\alpha}\frac{GM}{R^3}=\frac{2\sigma}{\alpha}\frac{\hbar^2}{m^2R^4}=\frac{\nu^4}{8\alpha\sigma^3}\frac{G^4M^4m^6}{\hbar^6}.
\end{eqnarray}
When $a\neq 0$, this expression  is asymptotically  valid for $R\rightarrow +\infty$ and $M\rightarrow 0$. (ii) In the TF approximation (for $a>0$), we get
\begin{eqnarray}
\label{pulse8}
\omega^2=\frac{2\nu GM}{\alpha R^3}=\frac{2\nu^{5/2}}{\alpha (6\pi\zeta)^{3/2}}
\frac{G^{5/2}Mm^{9/2}}{a^{3/2}\hbar^3}.
\end{eqnarray}
This expression is valid for $R\rightarrow R_{min}$, i.e. $M\rightarrow +\infty$, with $M(R)$ given by Eq. (\ref{mr5}). In that limit $\omega\rightarrow +\infty$. (iii) In the non-gravitational limit (for $a<0$), we find that
\begin{eqnarray}
\label{pulse9}
\omega^2=-\frac{2\sigma}{\alpha}\frac{\hbar^2}{m^2R^4}=-\frac{2\sigma^5}{\alpha(3\pi\zeta)^4}
\frac{m^2\hbar^2}{M^4a^4}.
\end{eqnarray}
This expression is valid for $R\rightarrow 0$ and $M\rightarrow 0$.

{\it Remark 1:} in the TF limit (when $a>0$), the density profile is known analytically, see Eq. (\ref{tf4}). Therefore, we can be more precise. If we use the Ledoux \cite{ledoux} formula $\omega^2=-2W/I$ for the pulsation (which is approximate) with the exact expressions (\ref{tf11}) and (\ref{tf7}) of $W$ and $I$, we obtain $\omega_{Ledoux}=0.3512 (GM/R_a^3)^{1/2}$.  The Gaussian approximation yields $\omega_{Gauss}=0.3199(GM/R_a^3)^{1/2}$ corresponding to Eq. (\ref{pulse8}). Finally, the exact result obtained by solving the Eddington \cite{eddingtonpuls} equation of pulsation numerically is $\omega_{exact}=0.3480(GM/R_a^3)^{1/2}$.  We note that the reference pulsation can be written $(GM/R_a^3)^{1/2}=N^{1/2}G^{5/4}m^{11/4}a^{-3/4}\hbar^{-3/2}$.

{\it Remark 2:} in the case of an attractive self-interaction ($a<0$), there is no equilibrium when $M>M_{max}$ and the system is expected to collapse. We could attempt to describe this collapse by using Eq. (\ref{vg3}) as done in Fig. \ref{collapseBEC}. For $R(t)\rightarrow 0$, it reduces to $\alpha M d^2R/dt^2=6\pi \zeta a\hbar^2 M^2/m^3R^{4}$ so that gravitational effects become negligible. This leads to a finite time collapse with a radius scaling like $R(t)\propto (t_{coll}-t)^{2/5}$. However, this scaling is different from the scaling $R(t)\propto (t_{coll}-t)^{1/2}$ obtained by directly solving the Gross-Pitaevskii equation (without gravity) \cite{sulem}. This shows that the Gaussian ansatz may not always be accurate. Eq. (\ref{vg3}) can be used close to a steady state so as to provide a good approximation of the pulsation period and of the growth rate, but it may lead to inaccurate results in more general situations.

\subsection{The radius versus scattering length relation}
\label{sec_rsl}

\begin{figure}[!h]
\begin{center}
\includegraphics[clip,scale=0.3]{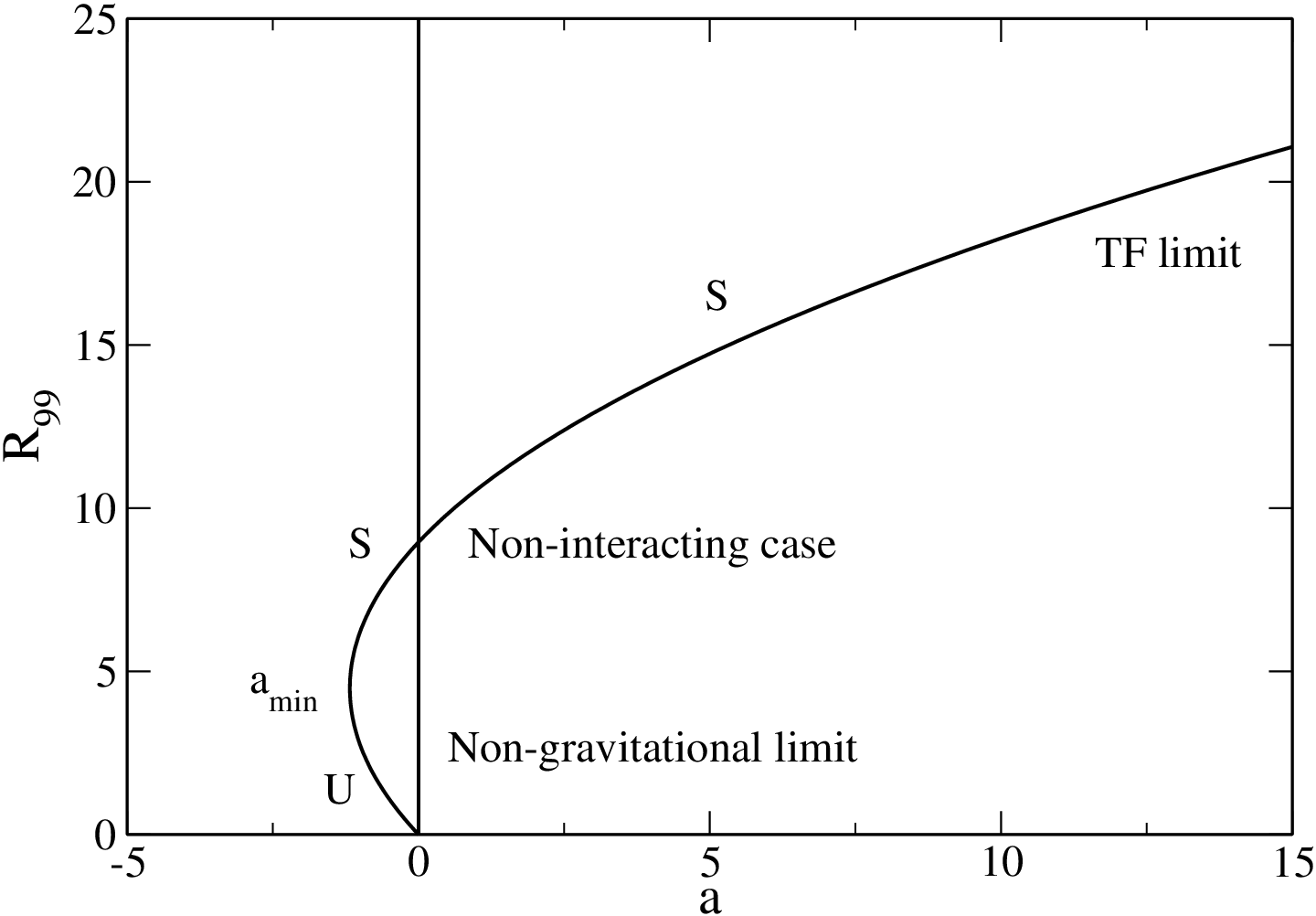}
\caption{Radius $R_{99}$ versus scattering length $a$ relation for a given value of the mass $M$. The radius is normalized by $R_Q$ and the scattering length by $a_Q$. Thus, $a=(\nu R^2-2\sigma R)/6\pi\zeta$ with $R_{99}=2.38167R$. The radius is given as a function of the scattering length by $R=(\sigma/\nu)(1\pm\sqrt{1+6\pi\zeta\nu a/\sigma^2})$ with the signs $+$ and $-$ for $a_{min}\le a\le 0$ and with the sign $+$ for $a\ge 0$. Stable solutions exist for $a\ge a_{min}=-\sigma^2/(6\pi\zeta\nu)= -1.178$ and $R_{99}\ge R_{99}^*$ corresponding to $R_*=\sigma/\nu$ i.e. $R_{99}^*= 4.477$. The radius $R$ is monotonically increasing with the scattering length $a$. The non-interacting limit ($a=0$) corresponds to $R=2\sigma/\nu$, i.e. $R_{99}= 8.955$. In the TF limit, valid for $a\rightarrow +\infty$, we have $R\sim (6\pi\zeta/\nu)^{1/2}a^{1/2}$, i.e. $R_{99}\sim 4.1252a^{1/2}$. The non-gravitational limit, valid for $R\rightarrow 0$ (lower branch) corresponds to $R\sim (3\pi\zeta/\sigma)|a|$, i.e. $R_{99}\sim 1.900|a|$, but these solutions are unstable.}
\label{Ra}
\end{center}
\end{figure}

\begin{figure}[!h]
\begin{center}
\includegraphics[clip,scale=0.3]{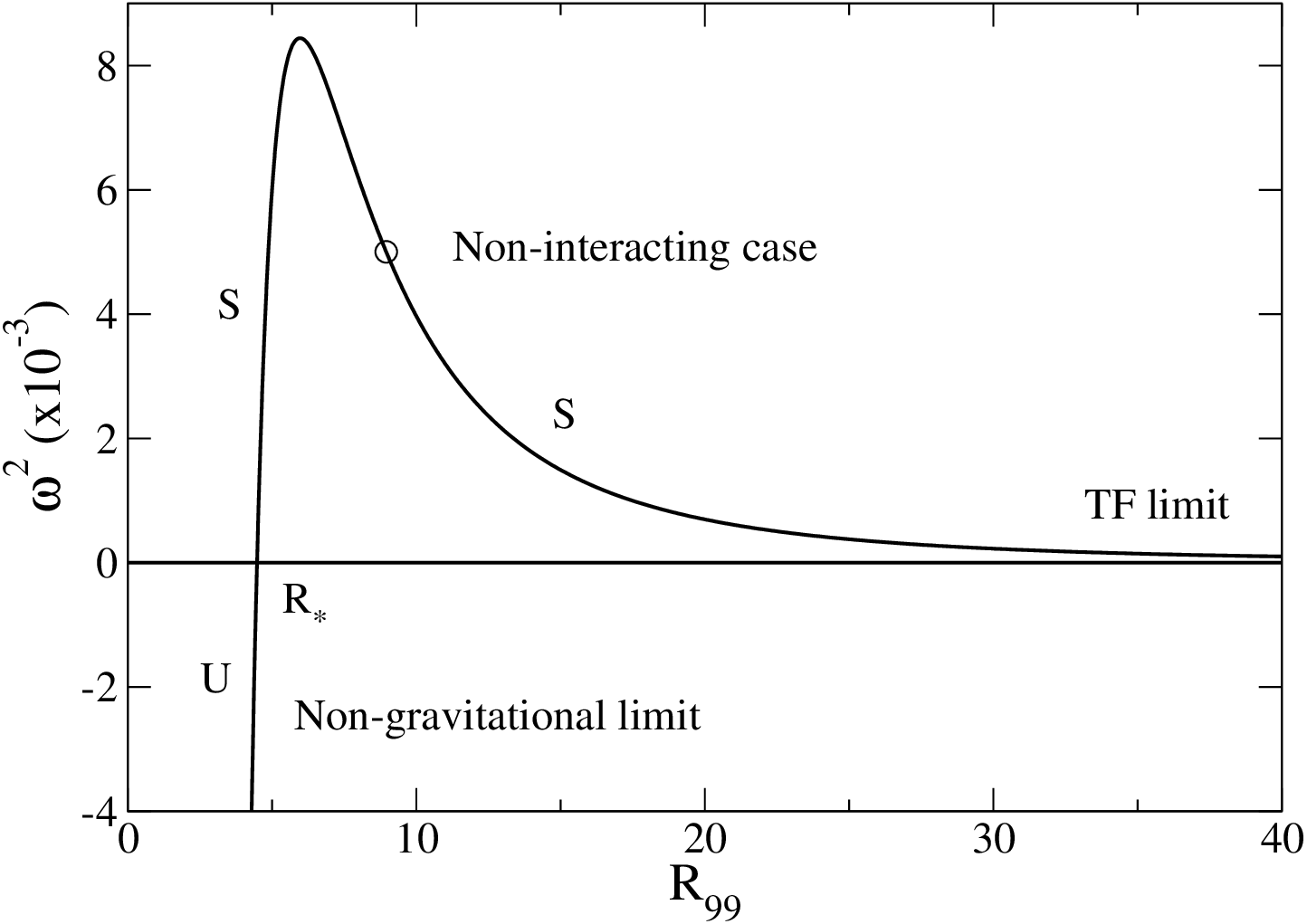}
\caption{Pulsation $\omega^2$ versus radius $R_{99}$ relation for a given value of the mass $M$. The pulsation is normalized by $\omega_Q=(GM/R_Q^3)^{1/2}=G^2M^2m^3/\hbar^3$ and the radius is normalized by $R_Q$. Thus, $\omega^2=(2\nu/\alpha R^3)(1-\sigma/\nu R)$ with $R_{99}=2.38167R$. In the non-interacting case ($a=0$), $\omega^2=\nu^4/8\alpha\sigma^3= 5.003\, 10^{-3}$. In the TF limit $R\rightarrow +\infty$,  $\omega^2\sim 2\nu/\alpha R^3\sim 7.186/R_{99}^{3}$. In the non-gravitational limit $R\rightarrow 0$, $\omega^2\sim -2\sigma/\alpha R^4\sim -32.18/R_{99}^4$. The pulsation is maximum for $R_{99}= 5.970$ with the value $\omega^2= 8.443\, 10^{-3}$. The pulsation can be expressed in terms of the scattering length as $\omega^2=\pm (2\nu^4/\alpha\sigma^3)\sqrt{1+6\pi\zeta\nu a/\sigma^2}/(1\pm \sqrt{1+6\pi\zeta\nu a/\sigma^2})^4$.}
\label{pulsation-MFIXE}
\end{center}
\end{figure}

For a given mass $M$, Eq. (\ref{fa5inv}) determines the radius $R$ as a function of the scattering length $a$. Inversely, Eq. (\ref{mr2}) yields
\begin{equation}
\label{rsl1}
a=\frac{m^3}{6\pi\zeta M\hbar^2}\left (\nu GMR^2-2\sigma\frac{\hbar^2}{m^2}R\right ).
\end{equation}
On the other hand, using Eqs. (\ref{pulse2}), (\ref{mr8}) and (\ref{rsl1}), the pulsation is given in terms of the radius by
\begin{equation}
\label{rsl2}
\omega^2=\frac{2\nu GM}{\alpha R^3}\left (1-\frac{\sigma\hbar^2}{\nu GMm^2R}\right ).
\end{equation}
The radius versus scattering length relation is represented in Fig. \ref{Ra} and the pulsation-radius relation is represented in Fig. \ref{pulsation-MFIXE}.

Let us first consider the case of repulsive short-range interactions ($a\ge 0$). There exists one, and only one, solution for any value of $a$ and it is stable ($\omega^2>0$). In the non-interacting case  $a= 0$, the radius is given by Eq. (\ref{mr3}) and the pulsation by Eq. (\ref{pulse7}). In the TF approximation, valid for $a\gg a_Q$, the radius is given by Eq. (\ref{mr4}) and the pulsation by Eq. (\ref{pulse8}). The radius increases with the scattering length $a$ and tends to $+\infty$ for $a\rightarrow +\infty$.   Therefore, a repulsive self-interaction allows to construct dark matter halos whose  size is much larger than for systems without self-interaction.

Let us now consider the case of attractive short-range interactions ($a<0$). There exists a minimum scattering length
\begin{eqnarray}
\label{rsl7}
a_{min}=-\frac{\sigma^2}{6\pi\zeta\nu}\frac{\hbar^2}{GM^2m},
\end{eqnarray}
corresponding to the radius
\begin{eqnarray}
\label{rsl8}
R_*=\frac{\sigma}{\nu}\frac{\hbar^2}{GMm^2}.
\end{eqnarray}
The approximate values $a_{min}=-1.178{\hbar^2}/{GM^2m}$ and $R_{99}^*=4.477{\hbar^2}/{GMm^2}$ obtained with the Gaussian ansatz are in fairly good agreement with the exact results $a_{min}^{exact}=-1.025{\hbar^2}/{GM^2m}$ and $(R_{99}^*)^{exact}=5.6{\hbar^2}/{GMm^2}$ obtained numerically in Paper II. For $a<a_{min}$, there is no equilibrium and the system undergoes gravitational collapse. For $a_{min}<a<0$, there  exists two solutions with different radii. The solution with the largest radius $R>R_*$ is stable ($\omega^2>0$) and the solution with the smallest radius  $R<R_*$ is unstable ($\omega^2<0$). In the stable region, the radius increases with the scattering length $a$. There exists a minimum radius $R_*$ given by Eq. (\ref{rsl8}). For $R\rightarrow R_*$, the pulsation behaves like $\omega^2=({2\nu GM}/{\alpha R_*^4})(R-R_*)$.
In the non-gravitational limit, corresponding to $R\ll R_*$, the radius is given by Eq. (\ref{mr6}) and the pulsation by Eq. (\ref{pulse9}).

\section{Jeans-type instability of a self-gravitating BEC}
\label{sec_jeans}

In this section, we study the linear dynamical stability of an
infinite homogeneous system of self-gravitating BECs described by the
quantum barotropic Euler equations. This is a generalization of the
classical Jeans problem to the fully quantum context. This type of
analysis has been performed by several authors
\cite{khlopov,bianchi,hu,silverman,sikivie,lee} in different
contexts. However, these authors did not take into account short-range
interactions that play an important role in the physics of cosmic
BECs.  Therefore, our discussion is more general. We shall also
develop the connection with the results established in the previous
sections.

Let us consider the quantum barotropic Euler-Poisson system
(\ref{mad3}), (\ref{mad7}) and (\ref{mfgp7}) describing
the dynamical evolution of an infinite system of self-gravitating
BECs. In order to correctly define the gravitational force in a {\it
static} infinite system (the case of an expanding universe is treated in \cite{cosmobec}), it is necessary to modify the Poisson
equation \cite{kiessling}. This can be done in different ways, either
by subtracting the average density $\overline{\rho}$ to the local
density $\rho({\bf r},t)$ or by introducing a shielding length
$\kappa^{-1}$ in the interaction and letting $\kappa\rightarrow
0$. This is not a ``Jeans swindle'' as oftentimes said \cite{bt} but
rather a well-defined and rigorous mathematical procedure
\cite{kiessling}. We therefore consider the system of equations
\begin{equation}
\label{j1}
\frac{\partial\rho}{\partial t}+\nabla\cdot (\rho {\bf u})=0,
\end{equation}
\begin{equation}
\label{j2}
\frac{\partial {\bf u}}{\partial t}+({\bf u}\cdot \nabla){\bf u}=-\frac{1}{\rho}\nabla p-\nabla\Phi+\frac{\hbar^2}{2m^2}\nabla\left (\frac{\Delta \sqrt{\rho}}{\sqrt{\rho}}\right ),
\end{equation}
\begin{equation}
\label{j3}
\Delta\Phi=4\pi G(\rho-\overline{\rho}),
\end{equation}
where $\overline{\rho}$ is the average density. In our study, the temperature $T=0$ and the pressure $p({\bf r},t)$ is due to short-range interactions (scattering) between particles. For specific applications, we shall assume a barotropic equation of state of the form (\ref{mad9}).

The linearized quantum barotropic Euler equations around an infinite homogeneous distribution with $\rho=\overline{\rho}$, ${\bf u}={\bf 0}$ and $\Phi=0$ are
\begin{equation}
\label{j4}
\frac{\partial\delta\rho}{\partial t}+\rho\nabla\cdot \delta{\bf u}=0,
\end{equation}
\begin{equation}
\label{j5}
\rho\frac{\partial \delta{\bf u}}{\partial t}=-c_s^2\nabla \delta\rho-\rho\nabla\delta\Phi+\frac{\hbar^2}{4m^2}\nabla (\Delta\delta\rho),
\end{equation}
\begin{equation}
\label{j6}
\Delta\delta\Phi=4\pi G \delta\rho,
\end{equation}
where $c_s^2=p'(\rho)=\rho h'(\rho)$ is the velocity of sound. For the equation of state (\ref{mad9}), it is given by
\begin{equation}
\label{j25}
c_s^2=\frac{4\pi a\hbar^2\rho}{m^3}.
\end{equation}
It is easy to combine these equations into a single equation for the perturbed density $\delta\rho$. We find that
\begin{equation}
\label{j8}
\frac{\partial^2\delta\rho}{\partial t^2}=-\frac{\hbar^2}{4m^2}\Delta^2\delta\rho+c_s^2\Delta \delta\rho+4\pi G\rho\delta\rho.
\end{equation}
Expanding the perturbation in plane waves of the form $\delta\rho({\bf r},t)\propto {\rm exp}\lbrack i({\bf k}\cdot{\bf r}-\omega t)\rbrack$, we obtain the dispersion relation
\begin{equation}
\label{j9}
\omega^2=\frac{\hbar^2 k^4}{4m^2}+c_s^2 k^2-4\pi G\rho.
\end{equation}
This is the gravitational analogue of the Bogoliubov \cite{bogoliubov}  energy spectrum of the excitation of a weakly interacting Bose-Einstein condensate. For large wavenumbers (small wavelengths), the quasi-particle energy tends to the kinetic energy of an individual gas particle and $\omega\sim \hbar k^2/2m$. The group velocity is
\begin{equation}
\label{j10}
v_g=\frac{\partial\omega}{\partial k}=\frac{\frac{\hbar^2 k^2}{2m^2}+c_s^2}{\sqrt{\frac{\hbar^2 k^4}{4m^2}+c_s^2k^2-4\pi G\rho}}k.
\end{equation}

In the non-interacting case ($a=0$), the pressure is zero ($p=0$) and the particles interact only via gravity. The dispersion relation reduces to
\begin{equation}
\label{j11}
\omega^2=\frac{\hbar^2 k^4}{4m^2}-4\pi G\rho.
\end{equation}
This equation exhibits a characteristic wavenumber due to the interplay between gravity and quantum effects (Heisenberg's uncertainty principle):
\begin{equation}
\label{j12}
k_J=\left (\frac{16\pi Gm^2\rho}{\hbar^2}\right )^{1/4}.
\end{equation}
This quantum Jeans scale appears in the works of \cite{khlopov,bianchi,hu,silverman,sikivie,lee}. The system is stable for perturbations with $k>k_J$ and unstable for perturbation with $k<k_J$. The maximum growth rate corresponds to $k=0$ leading to $\gamma=\sqrt{4\pi G\rho}$. The mass contained within the sphere of diameter $\lambda_J$, where $\lambda_J=2\pi/k_J$, is
\begin{equation}
\label{j13}
M_J=\frac{\pi}{6}\rho^{1/4}\left (\frac{\pi^3\hbar^2}{Gm^2}\right )^{3/4}.
\end{equation}
We therefore expect that the gravitational collapse of a homogeneous distribution of noninteracting bosons at $T=0$ leads to objects with typical radius $R_J=\lambda_J/2$ and typical mass $M_J$, or larger (recall that the maximum growth rate corresponds to $\lambda\rightarrow +\infty$) \footnote{The true mass and size of the structures depends on the evolution of the system in the nonlinear regime.}. The physical mechanism that leads to a non-vanishing Jeans scale and Jeans mass has the same nature as that which accounts for the equilibrium of the boson stars studied by Ruffini \& Bonazzola \cite{rb}. It corresponds to a balance between the gravitational force and the quantum pressure. Eliminating the density between Eqs. (\ref{j12}) and (\ref{j13}), we obtain the formula
\begin{equation}
\label{j14}
M_J=\frac{\pi^4}{12}\frac{\hbar^2}{Gm^2R_J},
\end{equation}
which qualitatively agrees with the mass-radius relation (\ref{nic2}) obtained in \cite{rb}.

In the Thomas-Fermi approximation in which the quantum potential can be neglected, the particles interact via gravity and they experience a pressure due to short-range interactions. The dispersion relation reduces to
\begin{equation}
\label{j15}
\omega^2=c_s^2 k^2-4\pi G\rho.
\end{equation}
This is the usual Jeans dispersion relation. For $a<0$, the system is always unstable. For $a>0$, the Jeans wavenumber is
\begin{equation}
\label{j16}
k_J=\frac{\sqrt{4\pi G\rho}}{c_s}=\left (\frac{Gm^3}{a\hbar^2}\right )^{1/2}.
\end{equation}
The system is stable for perturbations with $k>k_J$ and unstable for perturbation with $k<k_J$. The maximum growth rate corresponds to $k=0$ leading to $\gamma=\sqrt{4\pi G\rho}$.
The  characteristic wavenumber (\ref{j16}) arises due to the interplay between gravity and scattering.
We note that the Jeans wavenumber is independent on the density. The mass contained within the sphere of diameter $\lambda_J$, where $\lambda_J=2\pi/k_J$, is
\begin{equation}
\label{j18}
M_J=\frac{\pi}{6}\rho\left (\frac{\pi c_s^2}{G\rho}\right )^{3/2}=\frac{\pi}{6}\rho\left (\frac{4\pi^2a\hbar^2}{Gm^3}\right )^{3/2}.
\end{equation}
We therefore expect to form objects with typical radius $R_J=\lambda_J/2$ and typical mass $M_J$, or larger. The physical mechanism that leads to a nonvanishing Jeans scale and Jeans mass has the same nature as that which accounts for the equilibrium of self-gravitating BECs with repulsive short-range interactions studied by B\"ohmer \& Harko \cite{bohmer}. It corresponds to a balance between the gravitational force and the pressure due to repulsive scattering. In fact, Eq. (\ref{j16}) agrees with the radius (\ref{tf5}) of boson stars in the TF limit and Eq. (\ref{j18}) corresponds to the mass-central density relation (\ref{tf6}).

In the non-gravitational case, the dispersion relation reduces to
\begin{equation}
\label{j19}
\omega^2=\frac{\hbar^2k^4}{4m^2}+c_s^2 k^2.
\end{equation}
For $a>0$, the system is always stable. For $a<0$, the critical wavenumber is
\begin{equation}
\label{j20}
k_J=\left (\frac{16\pi |a|\rho}{m}\right )^{1/2}.
\end{equation}
The system is stable for perturbations with $k>k_J$ and unstable for perturbation with $k<k_J$ (the subscript $J$ - for Jeans - is here an abuse of language since gravity is neglected).
This  characteristic wavenumber arises due to the interplay between quantum pressure and attractive scattering.  The mass contained within the sphere of diameter $\lambda_J$, where $\lambda_J=2\pi/k_J$, is
\begin{equation}
\label{j21}
M_J=\frac{\pi}{6}\frac{1}{\rho^{1/2}}\left (\frac{\pi m}{4|a|}\right )^{3/2}.
\end{equation}
We therefore expect to form objects with typical radius $R_J=\lambda_J/2$ and typical mass $M_J$, or larger. Eliminating the density between Eqs. (\ref{j20}) and (\ref{j21}) we obtain
\begin{equation}
\label{j22}
M_J=\frac{\pi^2 m}{12|a|}R_J,
\end{equation}
which reproduces the scaling of Eqs. (\ref{dim6}) and (\ref{mr6}) \footnote{We have seen, however, that spatially inhomogeneous distributions with  $a<0$ are always unstable in the absence of gravitational forces. Therefore, the nonlinear development of the instability does {\it not} lead to equilibrium distributions in that case. We must therefore be careful in interpreting the formation of structures directly from the linear instability of a spatially homogeneous system.}. The growth rate is maximum for $k_*=({8\pi |a|\rho}/{m})^{1/2}$ and its value is $\gamma_*={4\pi|a|\hbar\rho}/{m^2}$.

\begin{figure}[!h]
\begin{center}
\includegraphics[clip,scale=0.3]{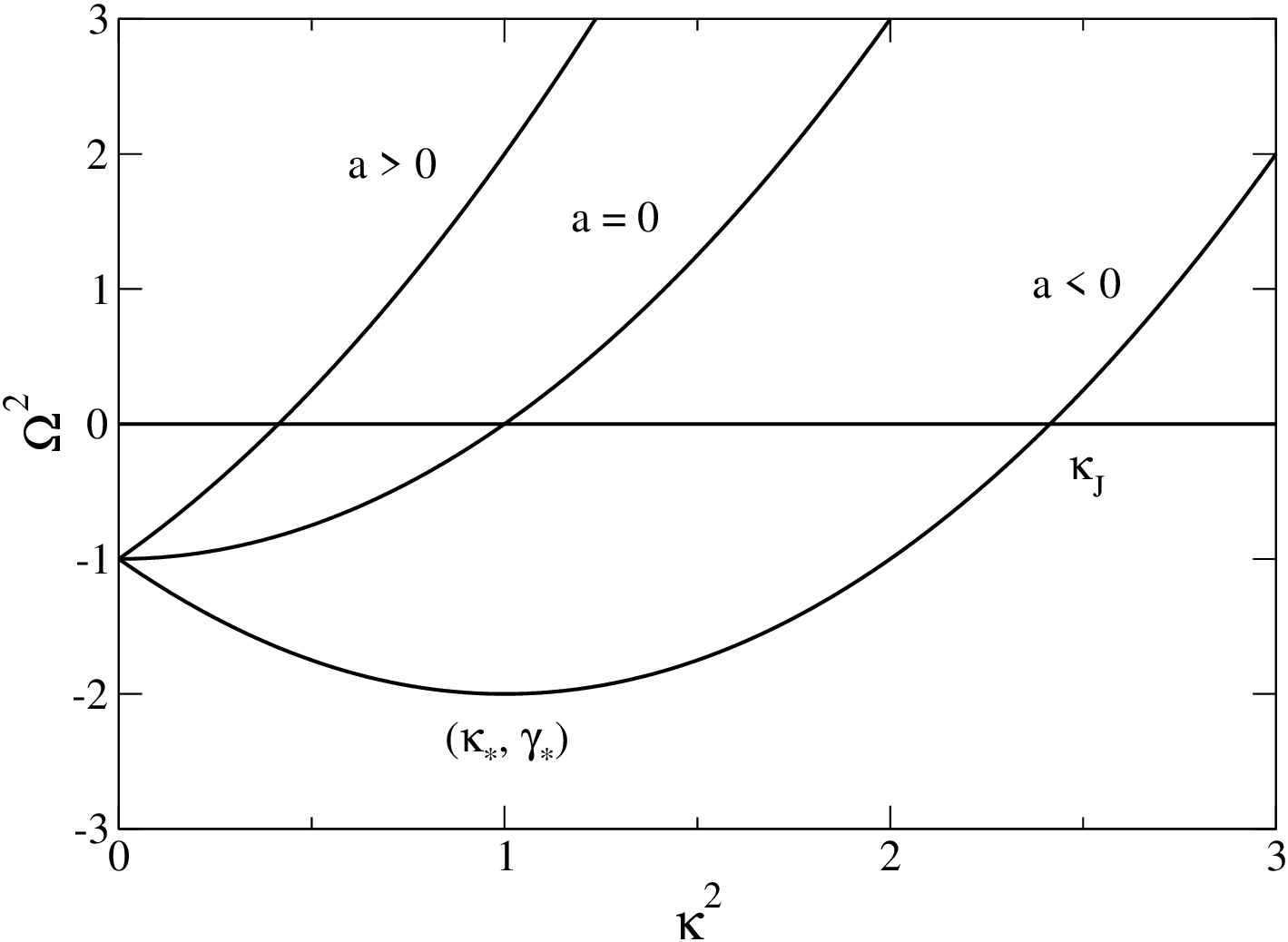}
\caption{Dimensionless dispersion relation $\Omega^2=\kappa^4+2\alpha\kappa^2-1$ with $\Omega=\omega/\omega_0$, $\kappa=k/k_0$ and $\alpha=a/a_0$ where $\omega_0=\sqrt{4\pi G\rho}$, $k_0=(16\pi G\rho m^2/\hbar^2)^{1/4}$ and $a_0=(Gm^4/4\pi\rho\hbar^2)^{1/2}$. The Jeans wavenumber is $\kappa_J^2=-\alpha+\sqrt{\alpha^2+1}$. For $a<0$, the maximum growth rate is $\gamma_*=\sqrt{1+\alpha^2}$ reached for $\kappa_*^2=-\alpha$. The TF limit corresponds to $\kappa^2\ll |\alpha|$, the non-gravitational limit to $\kappa^2\gg 1/|\alpha|$ and the non-interacting limit to $|\alpha|\ll 1$.}
\label{dispersion}
\end{center}
\end{figure}

From the dispersion relation (\ref{j9}), we note that the TF approximation is valid for $k\ll k_a$ with $k_a\equiv (16\pi |a|\rho/m)^{1/2}$ and the non-gravitational approximation is valid for $k\gg k_g$ with $k_g=(Gm^3/|a|\hbar^2)^{1/2}$. The non-interacting limit corresponds to $k_a\ll k_g$ i.e. $a\ll a_{0}=(Gm^4/16\pi\rho\hbar^2)^{1/2}$. We now consider the general case. The pulsation vanishes at the critical Jeans wavenumber
\begin{equation}
\label{j27}
k_J^2=\frac{2m^2}{\hbar^2}\left\lbrack \sqrt{c_s^4+\frac{4\pi G\hbar^2\rho}{m^2}}-c_s^2\right\rbrack.
\end{equation}
Using the expression (\ref{j25}) of the velocity of sound, this can be rewritten
\begin{equation}
\label{j28}
k_J^2=\frac{8\pi |a|\rho}{m}\left\lbrack \sqrt{1+\frac{Gm^4}{4\pi a^2\hbar^2\rho}}-{\rm sgn}(a)\right\rbrack.
\end{equation}
The system is stable for $k>k_J$ and unstable for $k<k_J$. From the Jeans wavenumber (\ref{j28}), we can define the Jeans radius $R_J=\lambda_J/2=\pi/k_J$ and the Jeans mass $M_J=\frac{4}{3}\pi R_J^3$. We find that
\begin{equation}
\label{j29}
M_J=\frac{\frac{4}{3}\pi \left (\frac{\pi m}{8a}\right )^{3/2}}{\rho^{1/2}\left\lbrack \sqrt{1+\frac{Gm^4}{4\pi a^2\hbar^2\rho}}-{\rm sgn}(a)\right\rbrack^{3/2}}.
\end{equation}
Eliminating the density between Eqs. (\ref{j28}) and (\ref{j29}), we obtain
\begin{eqnarray}
\label{j30}
M_J=\frac{\pi^4}{12}\frac{\frac{\hbar^2}{m^2}R_J}{GR_J^2-\frac{\pi^2 a\hbar^2}{m^3}}.
\end{eqnarray}
This expression returns the mass-radius relation (\ref{mr2}). For $a>0$, $\omega^2$ increases monotonically with $k$. Accordingly, the growth rate is maximum for $k=0$ (infinite wavelength) leading to $\gamma=\sqrt{4\pi G\rho}$. For $a<0$, $\omega^2$ achieves a minimum negative value at
\begin{eqnarray}
\label{j31}
k_*=\left (\frac{8\pi |a|\rho}{m}\right )^{1/2}.
\end{eqnarray}
Accordingly, the growth rate is maximum for $k=k_*$ and its value is
\begin{eqnarray}
\label{j32}
\gamma_*=\sqrt{\frac{16\pi^2a^2\hbar^2\rho^2}{m^4}+4\pi G\rho}.
\end{eqnarray}
For $|a|\gg (Gm^4/4\pi\hbar^2\rho)^{1/2}$, we find that $\gamma_*\sim
4\pi|a_s|\hbar\rho/m^2\gg \sqrt{4\pi G\rho}$. Therefore, an attractive
short-range interaction ($a<0$) increases the growth rate of the
Jeans instability. The pulsation is plotted as a function of the
wavenumber in Fig. \ref{dispersion} for positive and negative
scattering lengths.

These results may have profound cosmological implications. Dark matter
is usually described by hydrodynamical equations of the form
(\ref{j1})-(\ref{j3}) without the quantum potential ($Q=0$). In the
context of cold dark matter (CDM) models with vanishing temperature
and pressure ($T=p=0$), the usual Jeans analysis predicts that all
scales are unstable. Consequently, the Jeans scale and the Jeans mass
are zero ($\lambda_J=M_J=0$). This is the intrinsic reason why CDM
models generate cuspy dark matter halo profiles and an abundance of
low mass halos. However, these cusps and satellites are not seen in
observations \cite{observations,satellites}. As argued by several
authors \cite{hu,sikivie,lee}, these problems could be solved if the
dark matter of the universe is a self-gravitating BEC. In that case,
the wave properties of the dark matter can stabilize gravitational
collapse, providing halo cores and suppressing small-scale
structures. Indeed, if dark matter is a BEC, even at $T=0$ there
exists a non-zero Jeans length ($\lambda_J>0$).  Stability below the
Jeans scale is guaranteed by the Heisenberg uncertainty
principle. Therefore, there exists a minimum radius and a minimum mass
at which the system is stable. If the particles have a repulsive
self-coupling ($a>0$), stability results from the pressure arising
from the repulsive scattering
\cite{bohmer}. This non-thermal pressure indeed stabilizes the system
against gravitational collapse and leads to dark matter halos with a
flat core equivalent to $n=1$ polytropes (or other barotropic
structures depending on the form of the
self-interaction). Alternatively, if the particles have an attractive
self-coupling ($a<0$), generating a negative pressure, this
(anti)-pressure can enhance the Jeans instability and fasten the
formation of structures. The virtues of these results could be
combined by assuming that the sign of the scattering length $a$
changes in the course of the evolution. It could be negative in the
early universe to help with the formation of structures and become
positive once the structures start to form in order to prevent their
complete gravitational collapse. The mechanism of this change of sign
is, however, unknown so that this idea remains highly speculative. In
any case, the quantum pressure and the scattering pressure are
small-scale effects. The BEC model and the CDM model differ at small
scales only while they are indistinguishable at large scales
\cite{hu}. Therefore, at large scales of observational interest, we
recover the usual equations of CDM which have proven to be very
relevant.

\section{Conclusion}

In this paper, we have studied the structure and the stability of a self-gravitating BEC with  short-range interactions. We have connected the results of Ruffini \& Bonazzola \cite{rb} obtained in the absence of self-coupling to the results of B\"ohmer \& Harko \cite{bohmer} obtained for self-coupled BECs in the Thomas-Fermi approximation. We have also considered the case of attractive short-range interactions ($a<0$) and found the existence of a maximum mass above which no equilibrium state exists. These results have been obtained analytically by using a Gaussian ansatz and developing an analogy with a simple  mechanical problem. In Paper II, we shall compare our approximate analytical results with the exact results obtained by numerically solving the equation of hydrostatic equilibrium. Finally, in Paper III, we shall extend our analytical method to more general situations.

Our study was motivated by the proposal that dark matter halos could be gigantic cosmic BECs \cite{baldeschi,sin,hu,leekoh,peebles,goodman,arbey,bohmer}. In that case, gravitational collapse is prevented by the Heisenberg uncertainty principle or by the short-range interaction.  This suggestion remains, of course, highly speculative since we do not know the true nature of dark matter. There exists other theories according to which dark matter could be made of massive neutrinos \cite{fabbri,stella,gao,bilic1,bilic2,bilic3,bilic4}. In such theories, gravitational collapse is prevented by the Pauli exclusion principle for fermions. On the other hand, whatever the nature of its constituents,
if we view dark matter as a collisionless system described by the Vlasov equation, dark matter halos could result from a process of violent collisionless relaxation \cite{kull,csmnras}. In that  case, gravitational collapse is prevented by Lynden-Bell's type of exclusion principle \cite{lb}. This form of relaxation is much more rapid and efficient than a ``collisional'' relaxation (e.g. for fermions). Furthermore, it generates a density profile with a flat core (containing possibly a massive degenerate nucleus) and a $r^{-2}$ density halo yielding flat  rotation curves. These features are remarkably consistent with observations making this alternative scenario quite attractive.  It may also be recalled that the very existence of dark matter is questioned by some authors who introduced modified gravity theories, like the MOND theory \cite{milgrom}, to explain the astrophysical observations without invoking dark matter.

\appendix

\section{The velocity field}
\label{sec_velf}

In this Appendix, we determine the expression of the velocity field used in Sec. \ref{sec_efg} to compute the classical kinetic energy (\ref{efg6}).

The continuity equation (\ref{mad3}) can be rewritten
\begin{eqnarray}
\label{velf1}
\frac{\partial\ln\rho}{\partial t}+\nabla\cdot {\bf u}+\nabla\ln\rho\cdot {\bf u}=0.
\end{eqnarray}
Let us assume that $\rho({\bf r},t)$ is given by Eq. (\ref{efg1}) where $R=R(t)$ is a function of time. Then, we have
\begin{eqnarray}
\label{velf2}
\frac{\partial\ln\rho}{\partial t}=-3\frac{d\ln R}{dt}+2\frac{r^2}{R^2}\frac{d\ln R}{dt},
\end{eqnarray}
\begin{eqnarray}
\label{velf3}
\nabla\ln\rho=-3\frac{{\bf r}}{R^2}.
\end{eqnarray}
Assuming that the velocity profile is of the form  ${\bf u}({\bf r},t)=f(t){\bf r}$ and substituting these relations in Eq. (\ref{velf1}), we obtain
\begin{eqnarray}
\label{velf4}
\left (2\frac{r^2}{R^2}-3\right )\left \lbrack \frac{d\ln R}{dt}-f(t)\right \rbrack=0.
\end{eqnarray}
This relation is identically satisfied if $f(t)=d\ln R/dt$. Therefore, we can make the ansatz
\begin{eqnarray}
\label{velf5}
{\bf u}({\bf r},t)=\frac{d\ln R}{dt}{\bf r}=\frac{\dot R}{R}{\bf r},
\end{eqnarray}
for the velocity field.

\section{Relativistic extensions and maximum mass}
\label{sec_relat}

In this Appendix, we propose a simple way to generalize our Newtonian results to the relativistic regime. Our approach returns the right scaling of the maximum mass in the known cases and provides interpolation formulae in more general cases.

\subsection{Thomas-Fermi limit for fermion stars}
\label{sec_rf}

The mass-radius relation of a non-relativistic fermion star is
\begin{eqnarray}
\label{rf1}
R\sim \frac{\hbar^2}{GM^{1/3}m^{8/3}},
\end{eqnarray}
with a prefactor $4.51$ \cite{chandra}. According to this relation, there exists an equilibrium state for any value of the mass $M$. In reality, this relation ceases to be valid when the radius approaches the Schwarzschild radius $R_S={2GM}/{c^2}$ so that general relativistic effects come into play. Equating the two relationships, and introducing the Planck mass $M_P=({\hbar c}/{G})^{1/2}$, we obtain the maximum mass of a relativistic fermion star
\begin{eqnarray}
\label{rf4}
M_{Ch}\sim \frac{M_P^3}{m^2}.
\end{eqnarray}
This corresponds to the Chandrasekhar mass which scales like Eq. (\ref{rf4}) with a prefactor $0.376$ in general relativity \cite{ov}. The corresponding radius, which can be interpreted as the minimum radius of a relativistic fermion star is
\begin{eqnarray}
\label{rf5}
R_{min}\sim \frac{M_P}{m}\lambda_c,
\end{eqnarray}
where $\lambda_c={\hbar}/{mc}$ is the Compton wavelength. The prefactor in Eq. (\ref{rf5}) is $3.52$ \cite{ov}.  Since $m/M_P\ll 1$, we note that $R_{min}\gg \lambda_c$. We also note that ${\lambda_c}/{l_P}={M_P}/{m}$ where $l_P=({\hbar G}/{c^3})^{1/2}$
is the Planck length.

\subsection{Non-interacting boson stars}
\label{sec_rbn}

In the absence of short-range interaction, the mass-radius relation of a non-relativistic self-gravitating BEC is given by \cite{rb,membrado}:
\begin{eqnarray}
\label{rbn1}
R\sim \frac{\hbar^2}{GMm^2},
\end{eqnarray}
with a prefactor $9.9$ (if $R$ represents the radius containing $99\%$ of the mass). This relation is valid as long as the radius is much larger than the Schwarzschild radius $R_S={2GM}/{c^2}$. Equating the two relationships, and introducing the Planck mass, we obtain the maximum mass of a relativistic self-gravitating BEC without self-interaction
\begin{eqnarray}
\label{rbn2}
M_{Kaup}\sim \frac{M_P^2}{m}.
\end{eqnarray}
This corresponds to the Kaup \cite{kaup} mass which scales like Eq. (\ref{rbn2}) with a prefactor $0.633$. The corresponding radius, which can be interpreted as the minimum radius of a relativistic self-gravitating BEC without short-range interaction is
\begin{eqnarray}
\label{rbn3}
R_{min}\sim \lambda_c.
\end{eqnarray}
Therefore, the radius of a relativistic self-gravitating BEC without self-interaction scales like the Compton wavelength of the particles that compose the BEC. More precisely, the radius containing $95\%$ of the mass is given by Eq. (\ref{rbn3}) with a prefactor $6.03$ \cite{seidel90}. We note that the ratio between the classical radius (\ref{rbn1}) and the Compton wavelength scales like
\begin{eqnarray}
\label{rbn4}
\frac{R}{\lambda_c}\sim \frac{M_{Kaup}}{M},
\end{eqnarray}
with a prefactor $15.6$. Therefore, for non-relativistic BECs, such as dark matter galactic halos, the typical radius of the halo is much larger than the Compton wavelength ($R\gg \lambda_c$) since $M\ll M_{Kaup}$.

\subsection{Thomas-Fermi limit for boson stars}
\label{sec_rbo}

We now consider self-gravitating BECs with repulsive self-interaction ($a>0$). In the TF limit, the radius of the configuration is given by
\begin{eqnarray}
\label{rbo1}
R\sim\left (\frac{a\hbar^2}{Gm^3}\right )^{1/2},
\end{eqnarray}
with a prefactor $\pi$ \cite{bohmer}. It is independent on the total mass $M$ of the system. In order to make contact with studies that consider self-coupled particles interacting via a $\frac{1}{4}\lambda |\phi|^4$ potential \cite{colpi,leekoh,goodman,arbey}, we introduce the dimensionless parameter
\begin{eqnarray}
\label{rbo2}
\frac{\lambda}{8\pi}\equiv \frac{a}{\lambda_c}=\frac{a m c}{\hbar},
\end{eqnarray}
measuring the strength of the short-range interaction. For $a=10^6 \, {\rm fm}$ and $m=1.44 \, {\rm eV}/c^2$ we get  $\lambda/8\pi=0.01$ and for $a=1 \, {\rm fm}$ and $m=14 \, {\rm meV}/c^2$ we find that $\lambda/8\pi=10^{-10}$. In terms of this parameter, the radius (\ref{rbo1}) can be rewritten
\begin{eqnarray}
\label{rbo3}
R\sim \left (\frac{\lambda \hbar^3}{8\pi Gc}\right )^{1/2}\frac{1}{m^2},
\end{eqnarray}
with a prefactor $\pi$.  Interestingly, the same relation appears in the work of Arbey {\it et al.} \cite{arbey}  who consider a self-coupled charged scalar field  and obtain an equation of state $p=(\lambda\hbar^3/4m^4 c)\rho^2$ equivalent to Eq. (\ref{mad9}). The same equation of state is obtained by Colpi {\it et al.} \cite{colpi} at low densities while $p=\rho c^2/3$ at high densities (see Appendix \ref{sec_eoscolpi}). There exists therefore a close connection between a self-coupled BEC described by the GP equation and a self-coupled charged scalar field. Using Eq. (\ref{rbo2}), the parameter $\chi$ introduced in Sec. \ref{sec_dim} can be written  $\chi=(\lambda/8\pi) (M/M_P)^2$. The TF approximation is valid provided that $(\lambda/8\pi) (M/M_P)^2\gg 1$.

The Newtonian approximation is valid as long as the radius (\ref{rbo1}) is much larger than the Schwarzschild radius $R_S={2GM}/{c^2}$. Equating these two relationships, we obtain the maximum mass
\begin{eqnarray}
\label{rbo4b}
M\sim \frac{\hbar c^2\sqrt{a}}{(Gm)^{3/2}}.
\end{eqnarray}
Using Eq. (\ref{rbo2}) and introducing the Planck mass, it can be rewritten 
\begin{eqnarray}
\label{rbo4}
M\sim \sqrt{\lambda} \frac{M_P^3}{m^2}.
\end{eqnarray}
This is the maximum mass appearing in the work of Colpi {\it et al.} \cite{colpi} with a prefactor $0.062$.  As emphasized by these authors, for $\lambda\sim 1$, it scales like the Chandrasekhar mass ${M_P^3}/{m^2}$ while the Kaup mass scales like  ${M_P^2}/{m}$. In the presence of self-interaction, the maximum mass of a relativistic BEC  is much larger than the Kaup mass by a factor $M_P/m\gg 1$, so that it becomes astrophysically relevant. The radius of the configuration is still given, in the relativistic regime, by an equation of the form (\ref{rbo1}) or (\ref{rbo3}) but the prefactor is different. It can be written
\begin{eqnarray}
\label{rbo5}
R\sim\sqrt{\lambda}\frac{M_{P}}{m}\lambda_c,
\end{eqnarray}
with a prefactor $0.3836$ \cite{chavharko}. Therefore, for $\lambda\sim 1$, the radius of a self-coupled  BEC is much larger than the Compton wavelength since $M_P/m\gg 1$.

More generally, the mass (\ref{rbo4}) of a self-coupled BEC  is much larger than the Kaup mass (\ref{rbn2}) provided that $\lambda\gg (m/M_P)^2$. This is easily realized, even for weak self-interactions, since $m/M_P\ll 1$. Following Colpi {\it et al.} \cite{colpi}, it makes sense to introduce the parameter (we have not written their factor $4\pi$):
\begin{eqnarray}
\label{rbo6}
\Lambda=\lambda \frac{M_{P}^2}{m^2}.
\end{eqnarray}
Then, Eqs. (\ref{rbo4}) and (\ref{rbo5}) can be rewritten
\begin{eqnarray}
\label{rbo7}
M\sim \sqrt{\Lambda} \frac{M_P^2}{m}, \qquad R\sim \sqrt{\Lambda}\lambda_c,
\end{eqnarray}
with prefactors  $0.062$ and $0.3836$. The self-interaction is important for  $\Lambda\gg 1$ while it is negligible for $\Lambda\ll 1$. Even for small  $\lambda$, the self-interaction is important because $\Lambda$ can be quite large due to the greatness of the term $({M_P}/{m})^2$.

Finally, we emphasize that the radius $R$ of a self-coupled Newtonian
boson star (see Eqs. (\ref{rbo1}) and (\ref{rbo3})) depends on $m$ and
$a$ (or $\lambda$) only through the combination $a/m^3$ (or
$\lambda/m^4$). The same observation holds for the radius $R$ and the
mass $M$ of a self-coupled relativistic boson star (see Eqs. (\ref{rbo4b}), (\ref{rbo4}) and (\ref{rbo5})).

\subsection{Boson stars with attractive self-coupling}
\label{sec_att}

For self-gravitating BECs with attractive self-coupling ($\lambda<0$), using Eq. (\ref{rbo2}), the maximum mass (\ref{mr9}) and the minimum radius (\ref{mr10}) can be written
\begin{eqnarray}
\label{att1}
M\sim \frac{M_P}{\sqrt{|\lambda|}}, \qquad R\sim\sqrt{|\lambda|}\frac{M_{P}}{m}\lambda_c,
\end{eqnarray}
with prefactors $5.073$ and $1.1$ (if $R$ represents the radius containing $99\%$ of the mass).  For $|\lambda|\sim 1$, $M_{max}$ is of the order of the Planck mass $M_P= 2.18\, 10^{-8}\, {\rm kg}$, i.e. ridiculously small.  This essentially means that a self-gravitating boson gas with attractive interactions is very unstable. However, the mass increases when $|\lambda|\rightarrow 0$ while the radius decreases. The system becomes relativistic when the radius $R$ approaches the Schwarzschild radius $R_S={2GM}/{c^2}$ defined with the mass $M$. This happens when $|\lambda|< (m/M_P)^2$ yielding a maximum mass $M\sim M_P^2/m$, equivalent to the Kaup mass, and a radius $R\sim \lambda_c$ of the order of the Compton wavelength. Therefore, when self-coupling is attractive, the maximum mass is very low (implying a strong instability) except if the mass of the bosons, or their self-interaction, is extremely small (which is possible for axions and ultra-light bosons  \cite{mielke,mielkeschunck,hu}).

Summarizing, for $\lambda> (m/M_P)^2$ the maximum mass scales like $\sqrt{\lambda} M_P^3/m^2$ \cite{colpi}, for $\lambda=0$ it scales like $M_P^2/m$ \cite{kaup} and for $\lambda< -(m/M_P)^2$ it scales like $M_P/\sqrt{\lambda}$. It is interesting to obtain the three mass scales $M_P^3/m^2$, $M_P^2/m$, $M_P$ in the different regimes of a self-gravitating BEC.

\subsection{General case: an interpolation formula}
\label{sec_interpol}

In the general case, the mass-radius relation of a non-relativistic self-gravitating BEC with short-range interactions can be approximated by the relation (see Eq. (\ref{mr2})):
\begin{eqnarray}
\label{interpol1}
M\sim \frac{\frac{\hbar^2}{G m^2 R}}{1-\frac{a\hbar^2}{Gm^3R^2}},
\end{eqnarray}
where we have get rid of (uncertain) numerical factors. In the relativistic regime, equating the radius $R$  with the Schwarzschild radius $R_S={2GM}/{c^2}$ and using Eqs. (\ref{rbo2}) and (\ref{rbo6}), we obtain after simplification
\begin{eqnarray}
\label{interpol2}
M\sim \frac{M_P^2}{m}\sqrt{1+\Lambda},\qquad R\sim \lambda_c \sqrt{1+\Lambda}.
\end{eqnarray}
These formulae are valid for $\Lambda>-1$ (i.e.  $\lambda>-(m/M_P)^2$). For $\Lambda<-1$, the maximum mass and the minimum radius are given by the Newtonian expressions (\ref{att1}). In the non-interacting case $\Lambda=0$ and in the TF limit $\Lambda\rightarrow +\infty$, we recover the previous scalings (\ref{rbn2}), (\ref{rbn3}) and (\ref{rbo7}). Taking into account the correct prefactors obtained in \cite{kaup,colpi,seidel90,chavharko}, we can propose the interpolating formulae
\begin{eqnarray}
\label{interpol4}
M=0.633\sqrt{1+0.0096\Lambda}\frac{M_P^2}{m},
\end{eqnarray}
\begin{eqnarray}
\label{interpol5}
R_{95}=6.03  \sqrt{1+0.004\Lambda}\lambda_c,
\end{eqnarray}
for the maximum mass and the minimum radius of a relativistic self-gravitating BEC with short-range interactions. Interestingly, an expression for the critical mass of a relativistic boson star with repulsive self-interaction similar to Eq. (\ref{interpol4}) has been obtained in \cite{mielke} based on different arguments.

\section{Jeans instability with a cosmological constant}
\label{sec_cosmo}

In this Appendix, we extend the Jeans instability analysis of Sec. \ref{sec_jeans} by taking into account the cosmological constant $\Lambda= 3\, 10^{-56}\, {\rm cm}^{-2}$. This can be done in the Newtonian framework by replacing the Poisson equation by
\begin{eqnarray}
\label{cosmo1}
\Delta\Phi=4\pi G\rho-\frac{\Lambda c^2}{2}.
\end{eqnarray}
If we consider an infinite homogeneous distribution of matter that is solution of Eq. (\ref{cosmo1}), we find that the density is given by
\begin{eqnarray}
\label{cosmo2}
\rho=\frac{\Lambda c^2}{8\pi G}.
\end{eqnarray}
Let us assume that these particles are bosons forming Bose-Einstein condensates. We restrict ourselves to the non-interacting case ($a=0$) although more general situations could be contemplated as well. The Jeans analysis of Sec. \ref{sec_jeans}, combined with the expression (\ref{cosmo2}) of the density, shows that the gravitational collapse of these self-gravitating BECs is expected to form structures with typical radius and mass [see Eqs. (\ref{j12}) and (\ref{j13})] given by
\begin{eqnarray}
\label{cosm3}
R\sim \pi\left (\frac{\hbar^2}{\Lambda m^2 c^2}\right )^{1/4}, \quad M\sim \left (\frac{\Lambda c^2 \pi^8 \hbar^6}{G^4 m^6}\right )^{1/4}.
\end{eqnarray}
Now, using the expression of the cosmological constant in terms of classical fundamental constants proposed by B\"ohmer \& Harko \cite{bhcosmo}:
\begin{eqnarray}
\label{cosm4}
\Lambda=\frac{\hbar^2 G^2 m_e^6 c^6}{e^{12}},
\end{eqnarray}
and introducing the Planck mass, the Planck length and the fine structure constant $\alpha={e^2}/{\hbar c}= {1}/{137}$, we obtain
\begin{eqnarray}
\label{cosmo6}
R\sim \alpha^{3/2}\frac{M_P^2}{m^{1/2}m_e^{3/2}}l_P, \qquad M\sim\frac{1}{\alpha^{3/2}}\left (\frac{m_e}{m}\right )^{3/2}M_P.\nonumber\\
\end{eqnarray}
It is interesting to note that these length and mass scales can be expressed essentially in terms of fundamental constants.

\section{An approximate equation of state for relativistic BECs}
\label{sec_eoscolpi}

Colpi {\it et al.} \cite{colpi} model a boson star by a scalar field with a $\frac{1}{4}\lambda |\phi|^4$ interaction described by the Klein-Gordon-Einstein equations. In the Thomas-Fermi limit, they find that the scalar field becomes equivalent to a fluid with an equation of state
\begin{equation}
\label{eos1}
p=\frac{c^4}{36K}\left\lbrack \left (1+\frac{12K}{c^2}\rho\right )^{1/2}-1\right\rbrack^2,
\end{equation}
where $K=\lambda\hbar^3/4m^4 c$. For $\rho\rightarrow 0$ (small
densities), we recover the polytropic equation of state $p=K\rho^2$ of
a non-relativistic BEC with short-range interactions described by the
Gross-Pitaevskii equation (see Eqs. (\ref{mad9}) and
(\ref{rbo2})). For $\rho\rightarrow +\infty$ (high densities), we
obtain an equation of state $p=\rho c^2/3$ like in the core of
neutron stars. These asymptotic limits were not explicitly given in
\cite{colpi}. The study of BECs described by the 
Tolman-Oppenheimer-Volkoff equations with the equation of state
(\ref{eos1}) is considered in \cite{chavharko}. It is found that there
exists a maximum mass $M_{max}=0.06136\, \sqrt{\lambda}M_P^3/m^2$
(very close to the result obtained in \cite{colpi} by solving the
Klein-Gordon-Einstein equations) corresponding to a minimum radius
$R_{min}=0.3836 \, \sqrt{\lambda}(M_P/m)\lambda_c$ and a maximum
central density $(\rho_0)_{max}=1.592\, m^4c^3/\lambda
\hbar^3$.

\section{Conservation of the energy}
\label{sec_conservationE}

The energy associated with the GPP system (\ref{mfgp6})-(\ref{mfgp7}), or equivalently with the quantum barotropic Euler-Poisson system (\ref{mad3}), (\ref{mad7}) and (\ref{mfgp7}) is given by Eq. (\ref{ef1}). Let us explicitly show that it is conserved.  Using Eq. (\ref{ef3}), we have
\begin{eqnarray}
\label{cons1}
\delta \Theta_c=\int  \frac{{\bf u}^2}{2}\delta\rho\, d{\bf r}+\int \rho {\bf u}\cdot \delta {\bf u}\, d{\bf r}.
\end{eqnarray}
Using Eqs. (\ref{ef5}) and (\ref{mad5}), we find that
\begin{eqnarray}
\label{cons2}
\delta \Theta_Q=\frac{\hbar^2}{m^2}\int\nabla\sqrt{\rho}\cdot \delta\nabla\sqrt{\rho}\, d{\bf r}\nonumber\\
=\frac{\hbar^2}{m^2}\int\nabla\sqrt{\rho}\cdot \nabla\left (\frac{1}{2\sqrt{\rho}}\delta\rho\right )\, d{\bf r}\nonumber\\
=-\frac{\hbar^2}{2 m^2}\int \frac{\Delta\sqrt{\rho}}{\sqrt{\rho}}\delta\rho \, d{\bf r}=\frac{1}{m}\int Q \delta\rho \, d{\bf r}.
\end{eqnarray}
According to Eq. (\ref{ef6}), we have
\begin{eqnarray}
\label{cons3}
\delta U=\int H'(\rho)\delta\rho\, d{\bf r}=\int h(\rho)\delta\rho\, d{\bf r}.
\end{eqnarray}
Finally, for a symmetric binary potential of interaction
\begin{eqnarray}
\label{cons4}
\delta W=\int \Phi\delta\rho\, d{\bf r}.
\end{eqnarray}
Taking the time derivative of $E_{tot}$, using the previous relations, inserting the hydrodynamic equations (\ref{mad3}) and (\ref{mad6}), and integrating by parts, we easily obtain $\dot E_{tot}=0$. Note that this result remains valid if ${\bf u}$ is not a potential flow since ${\bf u}\cdot ({\bf u}\times (\nabla\times {\bf u}))={0}$.

\section{Virial theorem}
\label{sec_virialannexe}

In this Appendix, we establish the time-dependent virial theorem
associated with the quantum barotropic Euler-Poisson system
(\ref{mad3}), (\ref{mad7}) and (\ref{mfgp7}). For the sake of
generality, we derive it in $d$ dimensions. Taking the time derivative
of the moment of inertia (\ref{v2}) and using the continuity equation
(\ref{mad3}), we obtain after an integration by parts
\begin{eqnarray}
\label{virial2}
\dot I=2\int \rho {\bf r}\cdot {\bf u}\, d{\bf r}.
\end{eqnarray}
With the aid of the continuity equation (\ref{mad3}), the quantum barotropic Euler equation (\ref{mad7}) can be rewritten
\begin{eqnarray}
\label{virial3}
\frac{\partial}{\partial t}(\rho {\bf u})+\nabla(\rho {\bf u}\otimes {\bf u})\qquad\qquad\qquad\nonumber\\
=-\nabla p-\rho\nabla\Phi-\frac{\rho}{m}\nabla Q.
\end{eqnarray}
Taking the time derivative of Eq. (\ref{virial2}), substituting Eq. (\ref{virial3}), and using integrations by parts, we obtain the time-dependent virial theorem
\begin{eqnarray}
\label{virial4}
\frac{1}{2}\ddot I=2(\Theta_c+\Theta_Q)+d\int p\, d{\bf r}+W_{ii}.
\end{eqnarray}
For a steady state, $\ddot I=0$ and ${\bf u}={\bf 0}$, we obtain the equilibrium virial theorem
\begin{equation}
\label{virial9}
2\Theta_Q+d\int p\, d{\bf r}+W_{ii}=0.
\end{equation}
To obtain these expressions, we have used the following identities. First, we have introduced the virial of the gravitational force
\begin{equation}
\label{virial6}
W_{ii}=-\int \rho {\bf r}\cdot\nabla\Phi\, d{\bf r}.
\end{equation}
It can be shown that $W_{ii}=(d-2)W$ if $d\neq 2$ and $W_{ii}=-GM^2/2$ if $d=2$, where $W$ is the gravitational (potential) energy \cite{virial1}. 
On the other hand, it can be established after a few manipulations (using essentially integrations by part) that
\begin{equation}
\label{virial5}
-\int \frac{\rho}{m}{\bf r}\cdot \nabla Q\, d{\bf r}=2\Theta_Q.
\end{equation}
Indeed, using two successive integrations by parts, we have
\begin{eqnarray}
\label{virial10}
\int \rho {\bf r}\cdot \nabla \left (\frac{\Delta\sqrt{\rho}}{\sqrt{\rho}}\right )\, d{\bf r}=\nonumber\\
d\int (\nabla\sqrt{\rho})^2\, d{\bf r}+2\int\nabla({\bf r}\cdot \nabla\sqrt{\rho})\cdot\nabla\sqrt{\rho}\, d{\bf r}.
\end{eqnarray}
Using the convention of summation over repeated indices, the last integral can be rewritten
\begin{eqnarray}
\label{virial11}
\int\partial_i(x_j\partial_j\sqrt{\rho})\partial_i\sqrt{\rho}\, d{\bf r}\nonumber\\
=\int x_j\partial_i\sqrt{\rho}\partial_i\partial_j\sqrt{\rho}\, d{\bf r}
+\int (\nabla\sqrt{\rho})^2\, d{\bf r}\nonumber\\
=\frac{1}{2}\int x_j \partial_j(\nabla\sqrt{\rho})^2\, d{\bf r}+\int (\nabla\sqrt{\rho})^2\, d{\bf r}\nonumber\\
=\frac{2-d}{2}\int (\nabla\sqrt{\rho})^2\, d{\bf r}.
\end{eqnarray}
Substituting the last term of this expression in Eq. (\ref{virial10}), we obtain
\begin{eqnarray}
\label{virial12}
\int \rho {\bf r}\cdot \nabla \left (\frac{\Delta\sqrt{\rho}}{\sqrt{\rho}}\right )\, d{\bf r}=2\int (\nabla\sqrt{\rho})^2\, d{\bf r}.
\end{eqnarray}
Combining Eqs. (\ref{virial12}) and (\ref{ef5}), we obtain
Eq. (\ref{virial5}).

\end{document}